**Electrically tunable, two-photon interference from remote silicon-vacancy centers in industrial silicon carbide**

F.D. Hrunski [1], D. Scheller [1], M. Hollendonner [1], K. Ullerich [1], S.K. Parthasarathy [1,2], C. Fu [1], A. Pointner [1], W. Knolle [3], F. Kaiser [4,5], D.B.R. Dasari [6] and R. Nagy [1,*]

[1] Institute of Applied Quantum Technologies, Friedrich-Alexander-Universität Erlangen-Nürnberg, Erlangen, Germany
[2] Fraunhofer Institute of Integrated Systems and Device Technology (IISB), Erlangen, Germany
[3] Leibniz-Institut für Oberflächenmodifizierung e.V., Leipzig, Germany
[4] Quantum Materials, Luxembourg Institute of Science and Technology (LIST), 28 Avenue des Hauts Fourneaux, Belval, 4362 Esch-sur-Alzette, Luxembourg
[5] Department of Physics and Materials Science, University of Luxembourg, 2 Avenue de l'Université, Belval, 4365 Esch-sur-Alzette, Luxembourg
[6] 3. Physikalisches Institut, IQST and Centre for Applied Quantum Technologies, University of Stuttgart, Stuttgart, Germany

[*] roland.nagy@fau.de

*Distributed quantum networks rely on spatially separated, independently operated quantum systems as network nodes, whose emitted photons must be interfered with high visibility to establish end-to-end entanglement. Crucially, for network-relevant applications, high visibilities must be achieved over prolonged timescales to reduce overheads for error correction, and to increase network rates. Here, we demonstrate experimentally that silicon vacancy ($V_{Si}$) color centers in silicon carbide (SiC) achieve these requirements, notably in a mass-deployable fashion. We integrate $V_{Si}$ centers in different industrial-grade SiC p-i-n diodes, which are controlled via voltage biassing. This way, we demonstrate both, spectral overlapping of 19 randomly selected $V_{Si}$ centers in different diodes, as well as spectral narrowing close to the lifetime limit, i.e., typically below 60 MHz. Notably, these performance parameters are long-term stable, e.g., readjusting the p-i-n diode bias is required only every 8.4 hours, which reduces significantly the overall experimental overhead. We then use these assets to demonstrate high-quality two-photon interference between $V_{Si}$ centers located in two different cryostat setups, which are spatially separated by two meters. Notably, we perform a 26-days long measurement campaign, demonstrating two-photon interference with state-of-the-art raw interference visibilities of 82%, which aligns with the current state-of-the-art. These results establish $V_{Si}$ centers in industry-grade SiC devices as a scalable, spectrally stable building block for distributed quantum networks.*

Quantum networks create photon-mediated entanglement between separated quantum systems to enable new digital technologies, such as provably secure communication, distributed quantum computing, and enhanced sensing networks[1,2]. Color centers in solid-state hosts have emerged as promising candidates for a scalable deployment of such network nodes, combining a long-lived electron-spin memory with a network-compatible optical interface[3,4]. Today, color center experiments rival with traditional demonstrations based on atoms[5] and ions[6]: heralded entanglement was shown between two independently operated nitrogen-vacancy (NV) centers in diamond[7], extended to a three-node NV network enabling multipartite entanglement distribution and entanglement swapping[8]. Similarly, group-IV color centers in diamond have shown excellent progress. A recent work demonstrated a two-node network with silicon-vacancy (SiV) centers in diamond including photon distribution over a telecom fiber network[9], and another work demonstrated two-photon interference from two separated tin-vacancy (SnV) centers in diamond[10].

Realizing the full potential of such networks requires nodes that combine two properties simultaneously: high photon indistinguishability and spectral stability sustained well beyond a single measurement session. Meeting these requirements for solid-state systems poses two distinct challenges. First, independently created color centers typically exhibit spectrally broadened lines due to local charge noise, as well as substantial inhomogeneous distributions in their optical transition frequencies. For NV centers in diamond several tens of $\mathrm{GHz}$ have been reported[11,12]. In diamond, this issue is addressed by pre-selecting suitable centers within narrow optical transition lines, and within a narrow frequency range, followed by frequency tuning via locally deposited Stark shift tuning electrodes[13,14]. The second challenge concerns the long-term reliable operation. Once two centers are spectrally matched, the related optical resonances must remain stable over different time scales. Short-term spectral stability needs to be substantially below optical linewidth under standard experimental conditions to fundamentally enable photon interference. This requirement has proven challenging, especially when high-power off-resonant laser pulses are required for the operation of a color center[15]. Long-term spectral stability is equally important since it directly correlates with the overall experimental performance. Improved long-term stability avoids time-consuming re-calibration and tuning instances, which improves the duty-cycle[16].

Here, we demonstrate that all the above challenges can be simultaneously overcome using color centers in a standard industrial semiconductor material. Our platform uses silicon vacancy ($V_{Si}$) color centers in 4H silicon carbide (4H-SiC), which has long been considered a promising candidate for quantum networks. Several experiments showed that $V_{Si}$ centers combine favorable electron-spin coherence and optical properties with the maturity of 4H-SiC as an industrially established semiconductor[17–21]. A remaining question to be addressed is whether $V_{Si}$ centers can indeed combine high interference visibility with long-term spectral stability. An initial experiment proved short-term stability of the optical transitions via a single-emitter Hong-Ou-Mandel interference experiment[22]. In this work, we address the remaining key questions to validate $V_{Si}$ centers for quantum networks: long-term stability, as well as optical interference between multiple separated $V_{Si}$ centers. To answer these questions, $V_{Si}$ centers were integrated into 4H-SiC *p-i-n* diode structures. This enables two complementary effects: a DC Stark shift that tunes the emission frequency, and a charge-depletion effect that suppresses fluctuations of the local electrostatic environment and thereby narrows the optical linewidth[23–25]. This separates two tasks otherwise coupled during fabrication: producing high-quality color centers and adjusting their optical transition frequencies, with the former set by the material platform and the latter performed electrically after fabrication. Building on this, we demonstrate reliable linewidth narrowing below $60\ \mathrm{MHz}$ across a set of $19$ individually characterized centers without pre-selection. Two color centers were then chosen in different 4H-SiC *p-i-n* diodes for investigation in a Hong-Ou-Mandel (HOM) experiment. To our knowledge, we present the first comprehensive investigation of two such $V_{Si}$ centers in separate 4H-SiC samples, held in separate cryostats two meters apart, in a free-space Hong-Ou-Mandel experiment, providing a conclusive result to qualify $V_{Si}$ centers for quantum network applications.

## Results

### *Electrically programmable $V_{Si}$ centers*

The V2 $V_{Si}$ center in 4H-SiC is a point defect formed by a missing silicon atom, hosting a spin-$3/2$ ground state and acting as a single-photon emitter at its zero-phonon line (ZPL) near $916\ \mathrm{nm}$, with spin-state-dependent optical contrast (Fig. 1a)[17,26]. Like most solid-state color centers, however, $V_{Si}$ transitions are subject to the surrounding charge environment: static charge distributions around each defect shift its emission frequency, while charge fluctuations broaden its linewidth. As a consequence, independently created $V_{Si}$ centers generally differ both in emission frequency and in linewidth. To perform a meaningful HOM experiment with high-visibility two-photon interference between two such centers, it is therefore critical to control both the emission frequency and the linewidth of the used $V_{Si}$ centers. To achieve this control, we embed $V_{Si}$ centers into a *p-i-n* diode structure (Fig. 1b). For this purpose, we use an industrially available 4H-SiC wafer with an integrated lateral $p^{++}$–$i$–$n^{++}$ doping profile (see Supplementary note SI1), enclosing a $4.1\ \mu\mathrm{m}$ wide intrinsic region.

For the two-photon interference experiment, we address the $A_2$ transition, which shows a higher cyclicity than the $A_1$ transition and thus emits more photons[17]. Fig. 1c shows a representative resonant PLE spectrum of the $A_2$ transition, recorded on a single $V_{Si}$ center at a reverse bias of $12.5\ \mathrm{V}$, yielding a linewidth of $27.3\ \pm\ 0.6\ \mathrm{MHz}$. To identify suitable candidates for the interference experiment, we characterized $19$ $V_{Si}$ centers across two separate samples at the common target frequency of $\nu_{\mathrm{T}}= 327.23100\ \mathrm{THz}$ ($916.15\ \mathrm{nm}$) as shown in Fig. 1d. This indicates that every $V_{Si}$ center was characterized at a different reverse bias voltage. Evaluating the resulting linewidths (Fig. 1d), we find that the *p-i-n* diode structure narrows the optical transition below $60\ \mathrm{MHz}$ across all characterized centers. Furthermore, recording the $A_2$ transition frequency as a function of reverse bias (Fig. 1e), each center follows a linear dependence of $4.33\ \pm\ 0.15\ \mathrm{GHz/(MV/m)}$, with a slope determined by its position relative to the depletion zone[24]. Hence, every characterized center can be electrically tuned into a common target resonance frequency, demonstrating that spectral matching between independently created $V_{Si}$ centers is achievable. In principle, any of these characterized centers could be selected for the HOM experiment. The $V_{Si}$ centers selected for the interference experiment on each sample were chosen based on the highest photon count rate among all characterized $V_{Si}$ centers, in order to keep the measurement duration of the interference experiment practical, rather than out of necessity. Based on this criterion, we selected $V_{Si,1A}$ with a linewidth of $34.26\ \mathrm{MHz}$ on sample A and $V_{Si,2B}$ with a linewidth of $58.83\ \mathrm{MHz}$ on sample B at reverse bias voltage of common emission frequency $\nu_{\mathrm{T}}$. Both $V_{Si}$ centers exhibited a Fabry-Pérot cavity filtered $A_2$ emission of 65 cps, the highest observed among all characterized centers.

### ***Two-photon interference between remote $V_{Si}$ centers***

Both $V_{Si}$ center samples have been integrated into a separate closed-cycle cryostat, each cooling the sample to $4\ \mathrm{K}$ (Fig. 2a). Resonant photoluminescence excitation (PLE) spectroscopy is performed using a $916\ \mathrm{nm}$ laser, which is amplitude-modulated by an acousto-optic modulator (AOM) and by a phase-electro-optic modulator (EOM) to switch the excitation on and off and to simultaneously drive the $A_1$ and $A_2$ transitions in a dual-tone scheme.

For faster data acquisition over the extended measurement campaign presented below, the HOM measurement is performed under off-resonant rather than resonant excitation. The resulting linewidths are characterized in the following. Each $V_{Si}$ center is excited using a $728\ \mathrm{nm}$ laser at an optical power of $1\ \mathrm{mW}$ and at a maximum count rate shortly above the saturation power. Under this excitation, the cavity-filtered $A_2$ transition exhibits an optical linewidth of $95.7\ \mathrm{MHz}$ for $V_{Si,1A}$ and $202\ \mathrm{MHz}$ for $V_{Si,2B}$, as directly measured by scanning the cavity filter across the emission line (see Supplementary notes SI2 and SI3). Deconvolving the known filter response yields intrinsic linewidths of $29.62\ \mathrm{MHz}$ for $V_{Si,1A}$ and $126.75\ \mathrm{MHz}$ for $V_{Si,2B}$. Notably, the deconvolved linewidth of $V_{Si,1A}$ remains comparable to that obtained under resonant excitation ($34.3 \pm 0.9\ \mathrm{MHz}$, Fig. 1d), demonstrating that off-resonant excitation does not necessarily preclude near-resonant spectral purity for $V_{Si}$ centers, though this behavior is not uniform across all characterized centers. The resulting emission is spectrally separated into the zero-phonon line (ZPL) and the phonon sideband (PSB). Photons in the ZPL are further filtered using a Fabry-Pérot cavity filter to transmit only the $A_2$ photons of $V_{Si,1A}$ and $V_{Si,2B}$, respectively, corresponding to signal-to-total count ratios of $S_1/I_1 = 0.99$ and $S_2/I_2 = 0.99$ (see Supplementary note SI3). The filtered $A_2$ photon streams from both nodes are then overlapped on a free-space $50{:}50$ beam splitter and detected by superconducting nanowire single-photon detectors, with detection events time-tagged for correlation analysis. Detailed information about the experimental setup is shown in the supplementary online information (see Supplementary notes SI4.1 and SI4.2).

To confirm single-photon emission and quantify the signal purity of both nodes, we performed Hanbury Brown–Twiss (HBT) measurements under off-resonant excitation at $1\ \mathrm{mW}$, the same optical power used for the subsequent HOM measurement. For $V_{Si,1A}$, recorded on the PSB, we obtain background-corrected $g^{(2)}_{11,\mathrm{PSB}}(0) = 0.0930(5)$. For the same $V_{Si}$ center, recorded on the cavity-filtered $A_2$ ZPL emission, we obtain $g^{(2)}_{11,A_2}(0) = 0.03(4)$ without background correction, since cavity filtering suppresses the background to below $1\%$ (Fig. 2c). Both values lie well below the classical single-photon threshold of $0.5$, confirming single-photon emission for this node. Corresponding HBT measurement of the PSB emission of $V_{Si,2B}$ is provided in Supplementary note SI5.1. To model the measured HOM data, we follow the framework of ref.[27–32] as applied to remote solid-state emitters:

$$G^{(2)}_{12}(\tau) = c_1^2 g^{(2)}_{11}(\tau) + c_2^2 g^{(2)}_{22}(\tau) + 2c_1c_2 \cdot \{1 - \eta \cdot \frac{S_1S_2}{I_1I_2} \cdot g^{(1)}_{11}(\tau) \cdot g^{(1)}_{22}(\tau) \cdot \cos(\Delta\omega_0\tau) \cdot \exp\left(-\frac{\Gamma|\tau|}{2}\right) \exp\left(-\frac{\sigma^2 \cdot \tau^2}{2}\right)\} \qquad (1)$$

Here, $g^{(2)}_{ii}(\tau)$ denotes the intensity autocorrelation function (HBT) of source $i$, weighted by its relative intensity $c_i$. $|g^{(1)}_{ii}(\tau)|$ of its first-order coherence with homogeneous linewidth $\gamma_i$. $S_i/I_i$ the signal-to-total count ratio; and $\eta$ a phenomenological factor for residual interferometric imperfections. To account for the finite spectral width of the Fabry-Pérot cavity filter used to isolate the $A_2$ transition, we extend this framework by an additional damping term $\exp(-\Gamma|\tau|/2)$, where $\Gamma$ is the Lorentzian filtered linewidth (see Supplementary note SI6). Jitter in the instantaneous emission frequency ($\Delta\omega_0 = \omega_1 - \omega_2$) due to the independent spectral diffusion processes of both the defects was also included. This is determined by the mean spectral displacement ($\Delta\omega_0$) and their linewidths ($\sigma$). The resulting HOM interference is shown in Fig. 3a. With both emitters held at $\nu_\mathrm{T}$ and their polarizations aligned in parallel, the normalized cross-correlation between the beam splitter outputs drops to $G^{(2)}_{12,\parallel}(0) = 0.091(3)$ at zero delay, well below the classical limit of $0.5$, thus showing direct evidence of two-photon interference between indistinguishable photons originating from two spatially separated $V_{Si}$ centers. As a control, rotating the polarization of one input arm by $90°$ renders the photons distinguishable while leaving all other experimental conditions unchanged. The normalized correlation returns then to $G^{(2)}_{12,\perp}(0) = 0.506$ (Fig. 3b), in agreement with the classical expectation. Both datasets are fitted with Equation (1), yielding a raw interference visibility of $V_{\mathrm{HOM,raw}} = 0.820(6)$.

***Sustained spectral matching***

To maintain the $A_2$ transition frequency of both nodes within the target spectral window, the emission frequency of each $V_{Si}$ center was checked every 2 hours via a resonant PLE scan, implementing a threshold-triggered proportional feedback on the reverse bias voltage of the respective *p-i-n* diode (see Supplementary note SI7). Fig. 4a and 4b show the resulting $A_2$ emission frequency traces of $V_{Si,1A}$ and $V_{Si,2B}$, respectively, remaining bounded around the common target frequency $\nu_{\mathrm{T}}$ throughout the full 628-hour measurement campaign as a result of this active stabilization. Without correction, the emission frequency of either $V_{Si}$ center was found to drift by up to $\pm 40$ MHz on a timescale of hours. The feedback loop was therefore configured to compensate the reverse bias voltage whenever the $A_2$ transition frequency deviated by more than $\pm 20$ MHz from $\nu_{\mathrm{T}}$. A more detailed analysis of the resulting frequency stability is shown in Fig. 4c, which shows the mutual detuning between both centers, $\delta\nu = \nu_{\mathrm{A_2,1A}} - \nu_{\mathrm{A_2,2B}}$, over the full campaign. This distribution is centered at a mean detuning of $1.9 \pm 1.5$ MHz, which fixes $\Delta\omega_0$ in Equation (1), with a standard deviation of $15.5 \pm 1.5$ MHz, which fixes σ in Equation (1). The mean detuning indicates that during the HOM measurement both $V_{Si}$ centers were detuned by a fixed value which leads to an oscillation with a period of 526.3 ns. Given our measurement window of $\pm 100$ ns, this oscillation is not resolved in the data. Furthermore, even under 728 nm excitation the $V_{Si}$ center shows only a mean spectral drift of $15.5 \pm 1.5$ MHz which is negligible in comparison to other solid-state systems[12].

To independently assess the spectral contribution to the measured visibility, we evaluate the expected two-photon interference visibility directly from the measured detuning distribution shown in Fig. 4c, following

$$V'_{\mathrm{HOM}} = \int d\delta\nu \, P(\delta\nu) \frac{4\gamma_1\gamma_2}{(\gamma_1+\gamma_2)^2+4\delta\nu^2} \qquad (2)$$

Here, $\gamma_1$ and $\gamma_2$ denote the homogeneous linewidths of $V_{Si,1A}$ and $V_{Si,2B}$, respectively $\gamma_1 = 34.26$ MHz, $\gamma_2 = 58.83$ MHz, $\delta\nu$ the instantaneous detuning between both transitions, and $P(\delta\nu)$ denotes the measured probability density of this detuning, taken as the Gaussian distribution characterized in Fig. 4c ($\mu = 1.9$ MHz, $\sigma = 15.5$ MHz). Evaluating Equation (2) yields a mean visibility of $V_{\mathrm{HOM}}' = 0.848$, representing the interference visibility expected from spectral mismatch alone, in the absence of any further experimental imperfections. Multiplying this spectral contribution with the residual interferometric factor $\eta = 0.963$ (Fig. 3), yields a combined visibility of $0.848 \times 0.963 \approx 0.817$, in good agreement with the measured raw visibility of $0.820(6)$.

Having established that the residual visibility gap is well accounted by the measured spectral mismatch between the two nodes, we now turn to the practical stability of the electrical tuning itself. The frequency of active corrections further quantifies how stable each node remains on its own between interventions. $V_{Si,1A}$ required correction at a mean interval of $19.4\ \mathrm{h}$ (Fig. 4d), remaining considerably longer within tolerance than $V_{Si,2B}$ with a mean interval of $8.4\ \mathrm{h}$ (Fig. 4e). We attribute this discrepancy to differing thermal fluctuations and mechanical vibrations between the two independently operated cryostats.

## Discussion

These results answer the question raised at the outset of this work: silicon-vacancy centers in industrial 4H-SiC can combine high two-photon interference visibility with spectral stability sustained over practically relevant timescales, when embedded in electrically controlled *p-i-n* diode structures. The raw visibility of $0.820(6)$, obtained between two independently operated, remotely located $V_{Si}$ centers, compares favorably with the state of the art for remote solid-state emitters: it matches the visibility reported for remote tin-vacancy centers in diamond[10], exceeds recent results for remote semiconductor quantum dots[33], and was sustained here not over a single measurement session but across a $628$-hour campaign, with both centers remaining spectrally stable for $8.4$ and $19.4$ hours on average between successive interventions, a regime in which comparable systems typically have not been characterized.

The origin of the residual visibility drop can be traced quantitatively. Removing the cavity filter used to isolate the $A_2$ transition thereby exposing the interference to the full, unfiltered spectral diffusion of $126.8\ \mathrm{MHz}$ of $V_{Si,2B}$ rather than the effective linewidth transmitted by the $75.25\ \mathrm{MHz}$ filter is estimated to reduce the visibility by approximately $8\ \%$. Due to a linewidth of $29.6\ \mathrm{MHz}$ for $V_{Si,1A}$ this effect is negligible. In contrast, the mean detuning of $1.9\,\mathrm{MHz}$ between the two nodes contributes negligibly to the measured visibility as $< 0.01\,\%$, as simulated with Eq. 1. The dominant contribution to the residual visibility gap is instead the session-to-session spectral diffusion, characterized by a standard deviation of $\sigma = 15.5\,\mathrm{MHz}$ in the long-term detuning distribution (Fig. 4c); evaluating Eq. 2 under idealized conditions further quantifies the individual contributions to this gap: even in the absence of spectral diffusion $\sigma = 0$, the non-Fourier-limited linewidths of both centers alone limit the achievable visibility to $93.03\,\%$, a reduction of approximately $7\,\%$ relative to the ideal case of fully transform-limited emitters. Assuming instead fully Fourier-transform-limited linewidths and vanishing spectral diffusion $\sigma = 0$, Eq. 2 would result in unity visibility.

Taken together, these results demonstrate that embedding $V_{Si}$ centers into *p-i-n* diode structures provides a significant, simultaneous advantage across the three properties required for network-relevant nodes: long-term optical stability, high achievable two-photon interference visibility, and scalability across many independently fabricated emitters for the realization of a distributed quantum network.

## Methods

### *Experimental setup*

All experiments were performed at cryogenic temperatures $< 10\,\mathrm{K}$ in two different attodry800 cryostats. Two self-build confocal microscopes were used to excite the V2 centers optically and detect their zero-phonon line emission as well as the redshifted phonon side band. Initialization and continuous as well as pulsed repumping was performed via two distinct $728\,\mathrm{nm}$ diode laser (Toptica iBeam Smart CD-728-A0). A cavity tunable diode laser (TOPAS PLC pro) was used for resonant excitation in combination with an acousto-optic modulator (Gooch-Housego AOMO 3200-1113) and a frequency modulative phase-EOM to excite both optical transitions simultaneously. The cavity tunable diode laser's frequency was tracked and locked by a wavemeter (High Finesse WS7). For additional spin contrast, a microwave AC voltage at $70\,\mathrm{MHz}$ was applied through the copper wire in the vicinity of the color centers (Fig. 1b). Excitation and emission photons were filtered on both confocal microscopes by a tunable long-pass filter (Semrock TLP01-995). Confocal scanning was performed with a scanning mirror (Mad City Labs, Nano-MTA2X10). Excitation was focused on the sample in the cryostat (Attocube attoDry 800) with a high NA objective (Zeiss Epiplan-Neofluar $100\mathrm{x}$, $\mathrm{NA}\ 0.9$). At off resonant CW operation the according energy density in the focal spot corresponds then to $1.6\ \mathrm{x}\ 10^{-5}\ \mathrm{J/cm^3}$. Phonon side band and zero-phonon line emission were separated through an additional tunable long-pass filter (Semrock FF925-Di01). The polarization state of the zero-phonon line photons was aligned through waveplates ($\lambda/2$ and $\lambda/4$), filtered out subsequently through Glan-Thompson polarizers ($916\ \mathrm{nm}$ AR coated from Bernhard-Halle Nachfl. Optische Werkstätten) and guided through polarization-maintaining fibers. For both ZPL emissions from different setups, two separate free-space, voltage-controlled Fabry-Pérot cavities (Thorlabs FPQFA-8) were used to filter out the photons originating from $A_1$ or $A_2$ lines. Single $A_2$-line photons were interfering on a balanced free-space beam splitter (Edmund Optics, AR coated). Interference the $A_2$-line photons and photons from the phonon side band were detected by a superconducting single-photon nanowire detector (Single Quantum Eos CS SQ201). Corresponding electrical pulses from the detector were examined via time-tagging electronics (Quantum Machines OPX & Swabian Time Tagger Ultra). For remote voltage control of the samples' reverse bias voltages and the cavities' scanning voltages a four-channel-SMU was used (Rohde & Schwarz HMP4040). Voltage signals for a stabilized cavity-transmission operation were aligned through a self-programmed, intervened and on single-photon-count based PID controlling loop. Voltage control of scanning mirrors was performed via two distinct data acquisition and analog output devices (National Instruments NI-USB6343). Additional lifetime measurements (SI5.2) were performed through a combination of the upper mentioned acousto-optic modulator and an amplitude-EOM (Jenoptik AM905b). The RF-voltage modulation for creating $< 1\ \mathrm{ns}$ pulses was realized with a programmable pulse generator (PicoQuant PPG 512).

### ***Sample fabrication and device structure***

Both samples were prepared from industrially available, industrially grown c-plane 4H-SiC p-i-n structured wafers (Epitaxially grown wafer: semi-insulating / n-type; supplier; wafer diameter of $6\ \mathrm{inch}$; polytype orientation $(0001)$, off-axis angle of $4 \pm 0.5°$) from JXT Technology Co. Ltd. According to the manufacturer, the dopant concentrations are $N_{\mathrm{Al^+}} = 2 \cdot 10^{19}\ \mathrm{cm^{-3}}$, $N_{\mathrm{N^-}} = 2 \cdot 10^{14}\ \mathrm{cm^{-3}}$ and $N_{\mathrm{N^-}} = 1 \cdot 10^{18}\ \mathrm{cm^{-3}}$ for corresponding *p*, *i* and *n* layers with thicknesses of $d_p = 2.0\ \mathrm{\mu m} \pm 8\ \%$, $d_i = 4.1\mathrm{\mu m} \pm$ $8\ \%$ and $d_n = 350\ \mu\mathrm{m}$. Ohmic contacts at the $p^{++}$ and $n^{++}$ sides were fabricated by $\mathrm{Ni/Al}$ deposition and subsequent annealing at $800\ °\mathrm{C}$ for $3\ \mathrm{min}$ at the $p^{++}$ side and at $1000\ °\mathrm{C}$ for $5\ \mathrm{min}$ at the $n^{++}$ side. Additionally, for the $p^{++}$ contact, a composition of $53\ \mathrm{nm}\ \mathrm{Ni}$ and $43\ \mathrm{nm}\ \mathrm{Al}$ was used, while a $50\ \mathrm{nm}$ layer of $\mathrm{Ni}$ was applied for the $n^{++}$ contact. To create single color centers, the sample was electron-irradiated with an energy of $4\ \mathrm{MeV}$ at a flux of $1\ \mathrm{x}\ 10^{12}\ \mathrm{cm}^{-2}$, followed by annealing at $600\ °\mathrm{C}$ for $30\ \mathrm{min}$. To ensure a higher yield of emitted photons, a $158\ \mathrm{nm}$ thick $\mathrm{SiO_2}$ anti-reflective coating was deposited onto the samples surface via atomic layer deposition. The obtained $V_{Si}$ centers were located directly in the samples' crystal bulk without use of any nanophotonic structures.

**Data availability**

The data sets that support the findings of the work presented in the article and its supplementary information are available from the corresponding author upon request.

## Acknowledgements

F.D.H. and R.N. acknowledge support for the research of this work from European Union under the Key Digital Technologies Joint Undertaking (KDT JU); now Chips Joint Undertaking (Chips JU), in the projects ARCHIMEDES (grant nos. 16MEE0329 and 101112295) and MOSAIC (grant nos. 16MEE0494 and 101081238), and from the German Federal Ministry of Education and Research (BMBF, now BMFTR) in the projects INNOBAT (grant no. 03XP0492D) and QuaLiProM (grant no. 03XP0573C). F.K. acknowledges support by the Luxembourg National Research Fund (FNR) through both, the PEARL chair "AQuaTSiC" under grant agreement 17792569, as well as the project "SiCqurTech" under the national grant agreement 18253399. The “SiCqurTech” project is additionally funded within the European Union’s Horizon 2020 Research and Innovation Programme under grant agreement 101017733. F.K. received further support via the European Research Council for the project "Q-Chip” under grant agreement 101171067, as well as the Horizon Europe Programme for the Flagship project "QIA Phase 1" under grant agreement 101102140. Additionally, F.D.H and R.N. would like to thank F. Gannott and A. Gumann from the Max Planck Institute for the Science of Light for their support in sample preparation. Furthermore, we thank J. Freitag, E. Renner and M. Gloßner from the Institute of Microwaves and Photonics, H. Weber from the Chair of Applied Physics, C. Becher from the Quantum Optics Group at Saarland University, and S. Götzinger from the Max Planck Institute for the Science of Light for fruitful discussions.

## Author Contributions

Methodology: F.D.H., R.N.; Measurements: F.D.H.; Help with experimental apparatus: F.D.H., M.H., S.K.P., A.P.; Fabrication: D.S., W.K.; Sample preparation: D.S., F.D.H.; Writing, reviewing & editing: F.D.H., D.S., M.H., K.U., S.K.P., C.F., A.P., F.K., D.B.R.D., R.N.; Data analysis: F.D.H., D.B.R.D., R.N.; Theoretical support & analysis: F.D.H., D.S., M.H., K.U., S.K.P., C.F., A.P., F.K., D.B.R.D., R.N.

## Competing interests

The authors declare no competing interests.

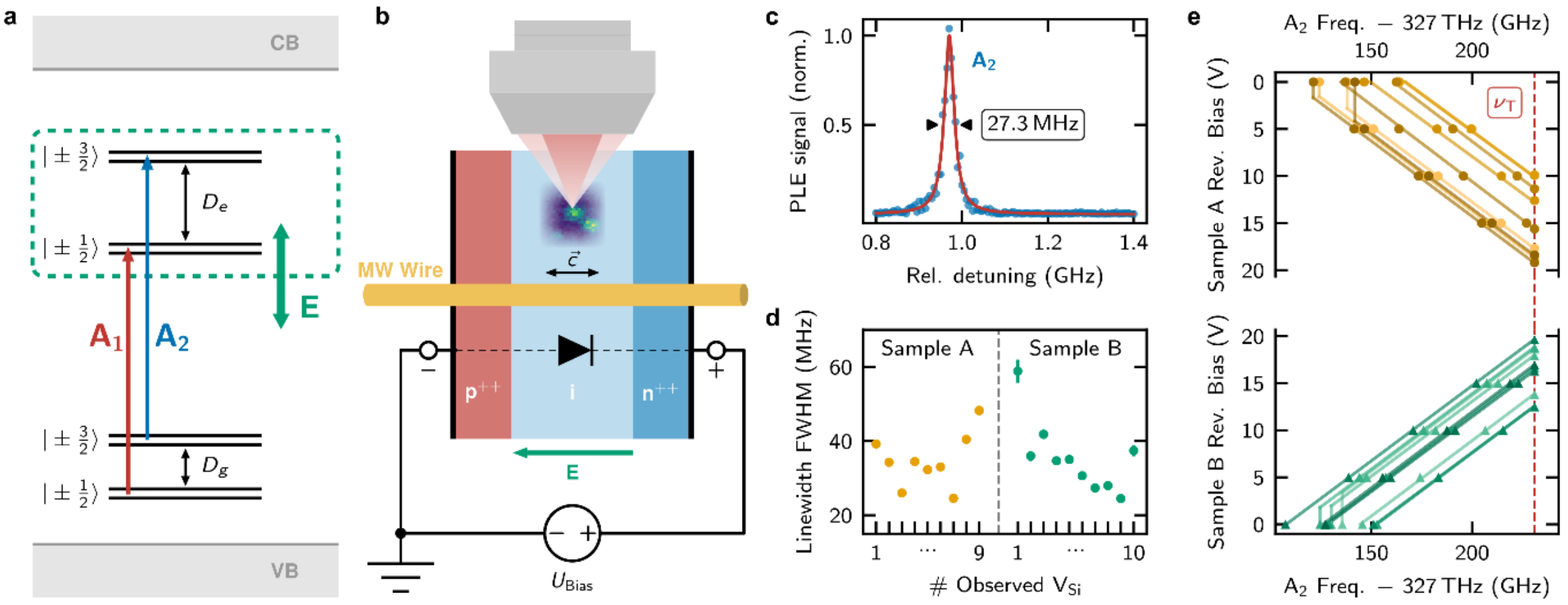

**Fig. 1| Single $V_{Si}$ centers in industrial 4H SiC *p-i-n* diode structures and their electrical tuning.**

**a** Level structure of the V2 $V_{Si}$ center. Ground and excited state are each split into $\pm 3/2$ and $\pm 1/2$ spin sublevels by zero-field splittings $D_g$ and $D_e$. The red and blue arrows denote the spin-conserving optical transitions $A_1$ and $A_2$, respectively. The green-dashed box marks the excited-state manifold, which is shifted by the applied electric field $E$ via the DC Stark effect. CB denotes the conduction band and VB the valence band. **b** Sample structure. Single silicon vacancy $V_{Si}$ centers are located in a $4.1\ \mu m$ wide intrinsic region of an industrially available lateral $p^{++}$- *i* - $n^{++}$ diode in industrially grown 4H-SiC and are addressed confocally. A reverse bias voltage $U_{Bias}$ is applied across the junction, and the resulting static electric field $E$ is oriented along the crystallographic $\vec{c}$ axis. A microwave (MW) wire enables MW-driven spin-state mixing. The inset shows a confocal photoluminescence scan of single $V_{Si}$ centers. **c** Dual-tone resonant photoluminescence excitation (PLE) spectrum of the $A_2$ transition of a representative center at a reverse bias of $12.5\ V$. Blue circles are the measured data and the red solid line is a Lorentzian fit, which yields a full width at half maximum (FWHM) of $27.3\ \pm\ 0.6\ MHz$. **d** Optical linewidths (FWHM) of the same $V_{Si}$ centers characterized in sample A (yellow circles) and sample B (green circles), extracted from Lorentzian fits to PLE spectra recorded at $\nu_T$. Error bars denote the standard error of the Lorentzian fit and are in part smaller than the marker size. **e** $A_2$ transition frequency as a function of reverse bias for nine V2-centers in sample A (yellowish circles) and ten centers in sample B (greenish triangles). Solid lines are linear fits. The red dashed line marks the common target frequency $\nu_T =\ 327.23100\ THz$.

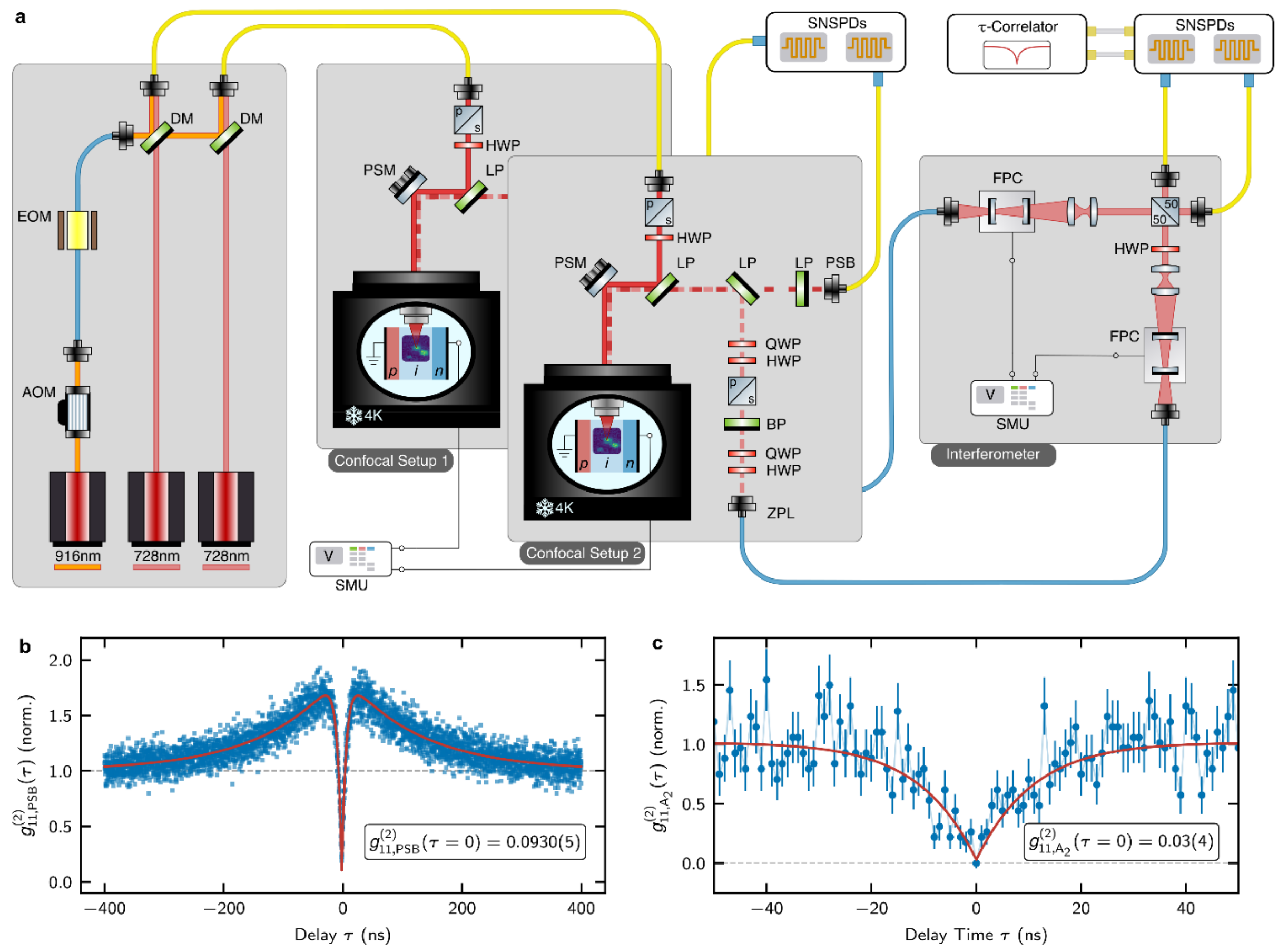

**Fig. 2| Setup for two-photon interference between two spatially separated $V_{Si}$ centers and single photon purity of the two emitters.**

**a** Experimental setup. Each $V_{Si}$ center is embedded in a *p*-*i*-n diode structure and hosted in one of two independently operated cryostats separated by two meters. Both centers are addressed in a confocal architecture, the zero-phonon-line (ZPL) emission is separated from the phonon sideband (PSB) by dichroic elements, and a single emission line ($A_2$) is selected from each ZPL by an individual Fabry-Pérot cavity before the two photon streams are interfered on a free-space 50: 50 beam splitter. Photons are detected by superconducting nanowire single-photon detectors and time-tagged for correlation analysis. Yellow lines indicate single-mode fibers and blue lines polarization-maintaining fibers. AOM, acousto-optic modulator; BP, optical bandpass filter; DM, dichroic mirror; EOM, electro-optic modulator; FPC, voltage-controlled Fabry-Pérot cavity; HWP, half-wave plate; LP, dichroic longpass; p/s Glan-Thompson-polarizers; PSM, piezoelectric fast scanning mirror; QWP, quarter-wave plate; SMU, remotely controlled voltage source; SNSPD, superconducting nanowire single-photon detector. **b** Background-corrected second-order autocorrelation of the PSB emission of $V_{Si,1A}$, measured in Hanbury Brown–Twiss (HBT) configuration. Blue squares are the normalized data and the red solid line is a fit to a three-level rate-equation model, which yields $g^{(2)}_{11,\mathrm{PSB}}(0) = 0.0930(5)$. **c** Corresponding autocorrelation function $g^{(2)}_{11,\mathrm{A}_2}(\tau)$ of $V_{Si,1A}$ recorded on cavity-filtered $A_2$ photons alone and without background correction, which yields $g^{(2)}_{11,\mathrm{A}_2}(0) = 0.03(4)$. Blue circles are the normalized data and the red solid line is a fit to the same model. Error bars denote Poissonian statistics.

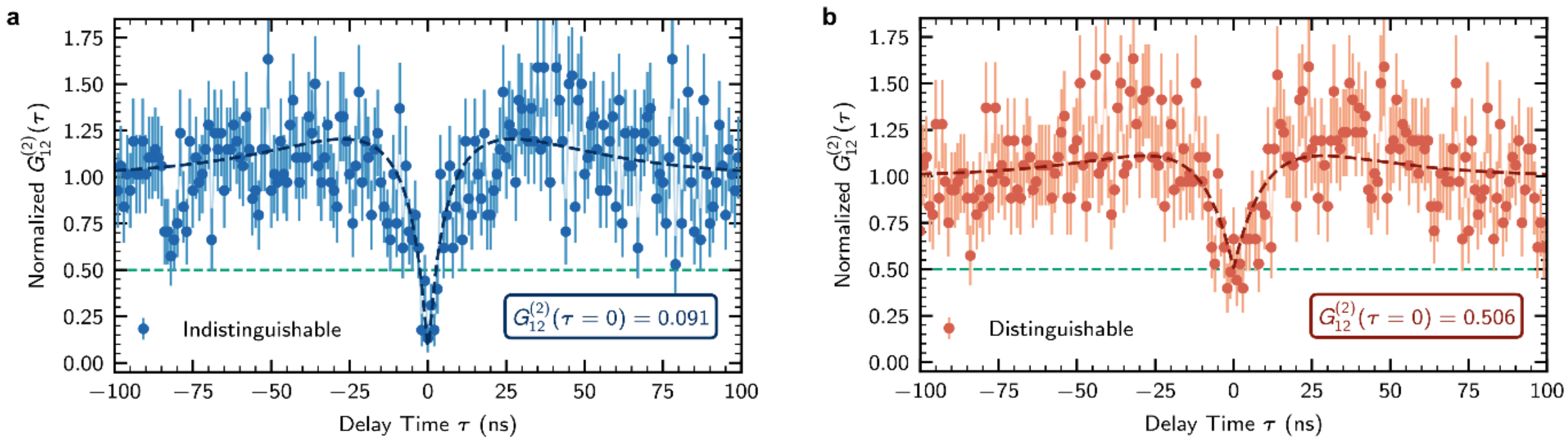

**Fig.3| Two-photon interference between remote $V_{Si}$ centers in two separate 4H-SiC samples.**

**a** Normalized cross-correlation $G^{(2)}_{12,\parallel}(\tau)$ of photons emitted by the two $V_{Si}$ centers with parallel photon polarizations. Blue circles are the measured data and the blue dashed line is a fit to the two-photon interference model in Equation (1). The green dashed line marks the classical two-photon limit of 0.5. The fit yields $G^{(2)}_{12,\parallel}(0) = 0.091(3)$. **b** Normalized cross-correlation $G^{(2)}_{12,\perp}(\tau)$ of the same two $V_{Si}$ centers with the polarization of one input arm before interference rotated by 90°, giving $G^{(2)}_{12,\perp}(0) = 0.506$. Red circles are the measured data and the red dashed line is a fit to the two-photon interference model in Equation (1). Error bars denote Poissonian counting statistics. Data were accumulated over 628 h per configuration in 1 ns bins.

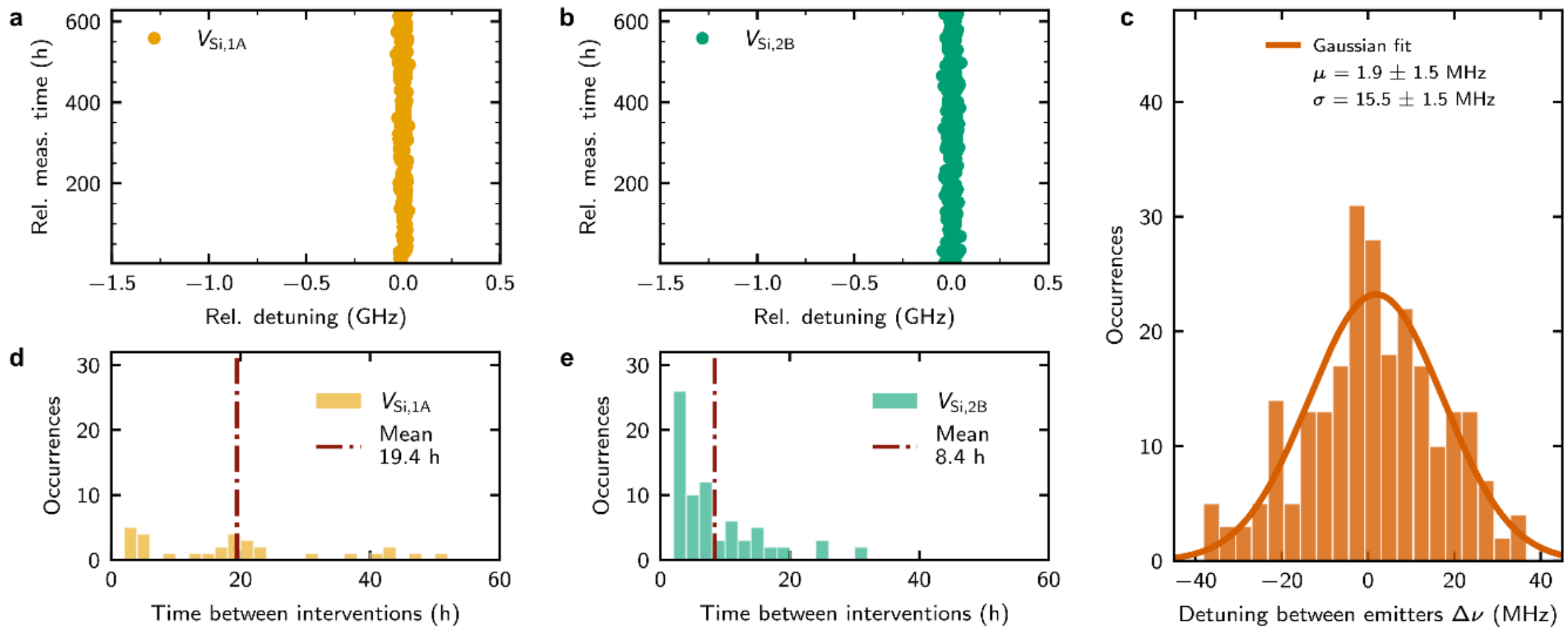

**Fig.4| Long-term spectral stability of interfered $V_{Si}$ centers and distribution of deduced two-photon interference visibility.**

**a** $A_2$ transition frequency of $V_{Si,1A}$ in sample A plotted relative to the common target frequency $\nu_T = 327.23100\,\mathrm{THz}$ over $628\,\mathrm{h}$ of cumulative measurement time. Each yellow circle is the center frequency obtained from a Lorentzian fit to one PLE scan. **b** Corresponding data for $V_{Si,2B}$ in sample B, shown as green circles. **c** Histogram of emission frequency detuning $\Delta\nu$ between the $A_2$ transitions of the two centers over the same period, obtained from the data in (**a**) and (**b**). The histogram has a bin width of $3.75\,\mathrm{MHz}$; the orange line is a Gaussian fit with mean $\mu = 1.9 \pm 1.5\,\mathrm{MHz}$ and standard deviation $\sigma = 15.5 \pm 1.5\,\mathrm{MHz}$. **d** Distribution of the time elapsed between successive feedback interventions on the diode bias voltage for $V_{Si,1A}$. Bin width is 2h. The dark red dash-dotted line marks the mean time elapsed between interventions, $19.4\,\mathrm{h}$. The longest interval without correction is $50\,\mathrm{h}$. **e** Corresponding distribution for $V_{Si,2B}$, with a mean interval between necessary interventions of $8.4\,\mathrm{h}$. The longest interval without correction is $32\,\mathrm{h}$.

## - Supplementary Information –

# Electrically tunable, two-photon interference from remote silicon-vacancy centers in industrial silicon carbide

F.D. Hrunski [1], D. Scheller [1], M. Hollendonner [1], K. Ullerich [1], S.K. Parthasarathy [1,2], C. Fu [1], A. Pointner [1], W. Knolle [3], F. Kaiser [4,5], D.B.R. Dasari [6] and R. Nagy [1]

[1] Institute of Applied Quantum Technologies, Friedrich-Alexander-Universität Erlangen-Nürnberg, Erlangen, Germany

[2] Fraunhofer Institute of Integrated Systems and Device Technology (IISB), Erlangen, Germany

[3] Leibniz-Institut für Oberflächenmodifizierung e.V., Leipzig, Germany

[4] Quantum Materials, Luxembourg Institute of Science and Technology (LIST), 28 Avenue des Hauts Fourneaux, Belval, 4362 Esch-sur-Alzette, Luxembourg
[5] Department of Physics and Materials Science, University of Luxembourg, 2 Avenue de l'Université, Belval, 4365 Esch-sur-Alzette, Luxembourg
[6] 3. Physikalisches Institut, IQST and Centre for Applied Quantum Technologies, University of Stuttgart, Stuttgart, Germany

## Contents

## SI1 Sample structure

Both samples were prepared from commercially available, industrially grown c-plane 4H-SiC *p-i-n* structured wafers (Epitaxially grown wafer: semi-insulating / n-type; supplier; wafer diameter of $6\,\mathrm{inch}$; polytype orientation $(0001)$, off-axis angle of $4 \pm 0.5°$) from JXT Technology Co. Ltd. According to the manufacturer, the dopant concentrations are $N_{\mathrm{Al}^+} = 2 \cdot 10^{19}\ \mathrm{cm}^{-3}$, $N_{\mathrm{N}^-} = 2 \cdot 10^{14}\ \mathrm{cm}^{-3}$ and $N_{\mathrm{N}^-} = 1 \cdot 10^{18}\ \mathrm{cm}^{-3}$ for corresponding *p*, *i* and *n* layers with thicknesses of $d_p = 2.0\ \mathrm{\mu m} \pm 8\,\%$, $d_i = 4.1\ \mathrm{\mu m} \pm 8\,\%$ and $d_n = 350\ \mathrm{\mu m}$ (Fig.SI1a). According to previous publication, we measured a dopant concentration of $N_{\mathrm{N}^-} = 9 \cdot 10^{14}\ \mathrm{cm}^{-3}$ in the intrinsic area[1]. Ohmic contacts at the *p*$^{++}$ and *n*$^{++}$ sides were made by $\mathrm{Ni/Al}$ deposition and subsequent annealing at $800\ °\mathrm{C}$ for $3\ \mathrm{min}$ at the *p*$^{++}$ side and at $1000\ °\mathrm{C}$ for $5\ \mathrm{min}$ at the *n*$^{++}$ side. Additionally, for the *p*$^{++}$ contact, a composition of $53\ \mathrm{nm\ Ni}$ and $43\ \mathrm{nm\ Al}$ was used, while a $50\ \mathrm{nm}$ layer of $\mathrm{Ni}$ was applied for the *n*$^{++}$ contact. To create single color centers, the sample was electron-irradiated with an energy of $4\ \mathrm{MeV}$ at a flux of $1\ \mathrm{x}\ 10^{12}\ \mathrm{cm}^{-2}$, followed by annealing at $600\ °\mathrm{C}$ for $30\ \mathrm{min}$. To ensure a higher yield of emitted photons (SI3), a $158\ \mathrm{nm}$ thick $\mathrm{SiO_2}$ anti-reflective coating optimized for $916\ \mathrm{nm}$ was deposited onto the samples surface via atomic layer deposition at $70\ °\mathrm{C}$. The obtained $V_{Si}$ centers were located directly in the samples' crystal bulk without use of any nanophotonic structures. During electrical operation, the *p*-side was grounded and the *n*-side was connected to a remote controllable voltage source, applying therefore the reverse bias voltage $U_{\mathrm{Bias}}$ (Fig. SI1b). The diode characteristics were verified at a cryogenic operating temperature of $5.2\ \mathrm{K}$ (Fig. SI1c).

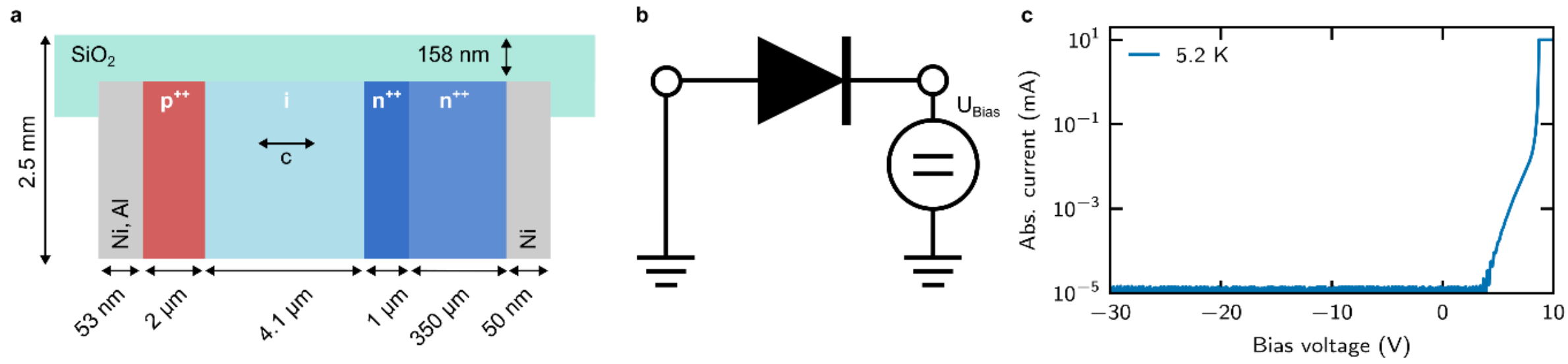

**Fig. SI1 Sample geometry, exemplary wiring and corresponding IV-characteristic.**

**a** Geometric structure and doping layers of manufactured *p-i-n* structured samples. Dopant concentrations are given in text. $c$ denotes the crystallographic axis. The scheme is not in scale. **b** Exemplary wiring scheme used to apply a reverse bias voltage $U_{\mathrm{Bias}}$ for tuning the $V_{Si}$ emission frequency via the DC Stark shift. **c** IV-characteristics of the sample at cryogenic temperature of $5.2\ \mathrm{K}$. For this measurement the bias was applied on the *p*-side, so the diode was operated in forward direction as is conventional for IV-characterization.

## SI2 Fabry-Pérot cavities for spectral narrow-band filtration

All off-resonantly excited spectroscopic measurements in this work were performed through the Fabry-Pérot filter cavities described in this section. The zero-phonon line (ZPL) of the V2 $V_{Si}$ center comprises two spin-conserving transitions, $A_1$ and $A_2$, separated by less than $1\ \mathrm{GHz}$ (Fig. 1a). Only $A_2$ photons were used for the two-photon interference experiment. The $A_1$ emission and the residual broadband background transmitted by the dichroic elements are therefore not sufficiently suppressed by these dichroics alone and must be filtered separately at each node. For this purpose, the ZPL emission of each node was passed through an individual free-space, piezo-tunable Fabry-Pérot cavity (Thorlabs FPQFA-8), resulting in two independently operated filter cavities for the two nodes A and B.

Both cavities were characterized individually before the interference experiment. This characterization serves two purposes. First, each cavity is used throughout this work as a scanning spectrometer. The spectra are recorded by sweeping the piezo voltage across the ZPL emission, which requires converting the applied voltage into an absolute optical frequency. Second, a measured spectrum is the convolution of the true emission line with the cavity transmission profile, recovering the intrinsic emitter linewidth (Eq. SI3) and modeling the coherence of the transmitted photons (SI6) therefore both require independent knowledge of the cavity's transmission bandwidth. Both quantities the voltage-to-frequency conversion and the transmission bandwidth were obtained from the same measurement. A resonant, narrow-linewidth ($<100\ \mathrm{kHz}$) tunable diode laser (TOPAS PLC pro), referenced to a wavemeter (High Finesse WS7), was used to probe each cavity. The piezo voltage was scanned four times in succession, with the laser fixed at a different frequency for each scan, the four reference frequencies were spaced by $5\ \mathrm{GHz}$. No modulation was applied. Each scan yields a single transmission maximum, whose position shifts linearly with the applied cavity voltage (Fig. SI2a, b). This linear relation gives voltage-to-frequency translation factors of $-1.8745\ \mathrm{GHz/V}$ for the cavity of node A and $-1.7380\ \mathrm{GHz/V}$ for the cavity of node B. The negative sign reflects the inverted response of the piezo actuators. Using these factors, the transmission profiles recorded in the same measurements were converted from volts to frequency and fitted with Lorentzian functions (Fig. SI2c, d). This yields transmission bandwidths (FWHM) of $\Gamma_\mathrm{A} = 66.08\ \mathrm{MHz}$ for node A and $\Gamma_\mathrm{B} =$ 75.25 MHz for node B, corresponding to finesse values of 454 and 399, respectively, and consistent with the manufacturer's specification.

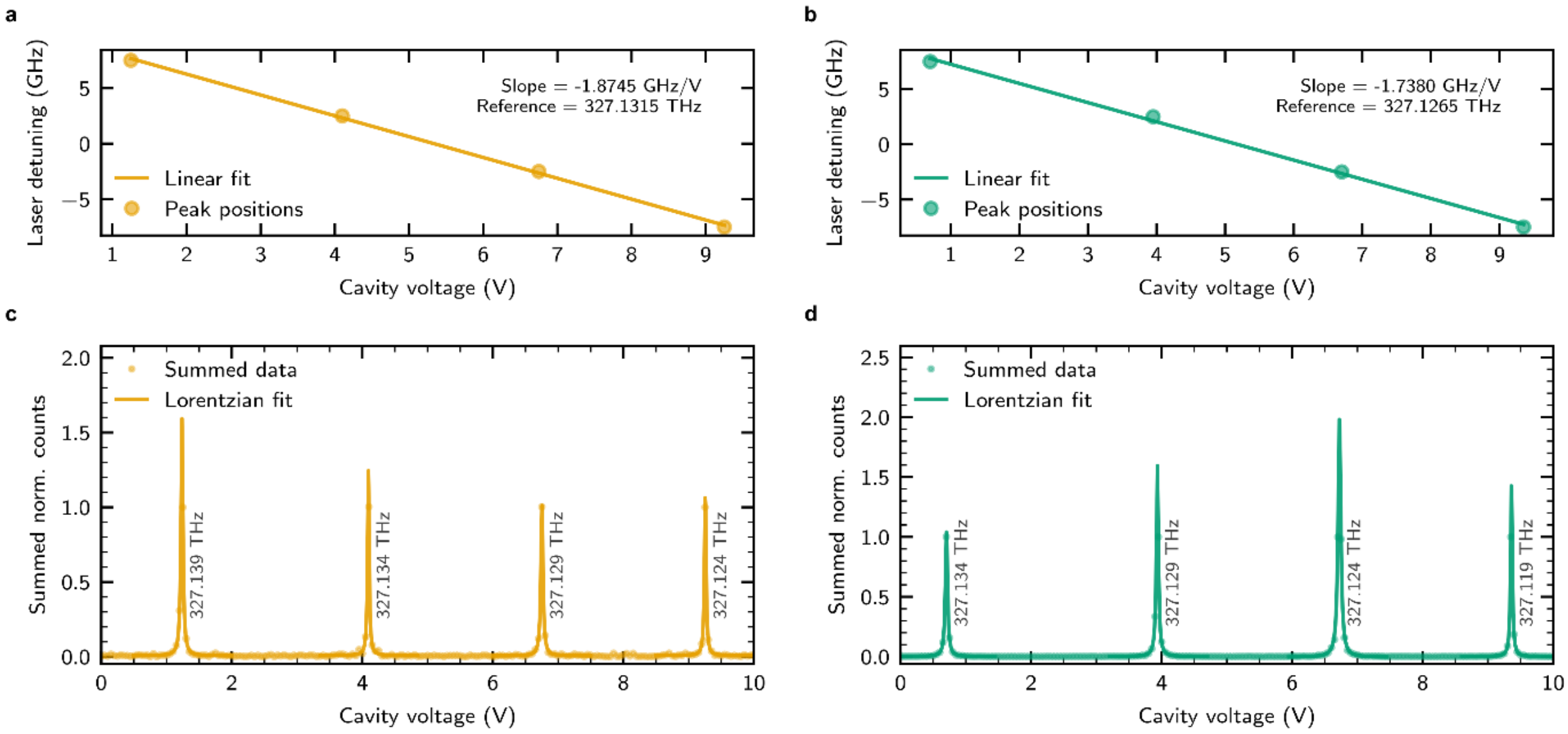

**Fig. SI2 Fabry-Pérot cavity characterization.**

**a** Voltage-to-frequency translation measurement of node's A filter cavity. Laser frequencies were fixed at values of 327.124000 THz, 327.129000 THz, 327.134000 THz and 327.139000 THz. For these values the transmission maximum was detected at different applied cavity voltages. Linear fit derives a translation factor of $-1.8745$ GHz / V for filter cavity of node A. **b** Corresponding measurement for node B cavity. Laser frequencies were 327.119000 THz, 327.124000 THz, 327.129000 THz and 327.134000 THz. Linear fit derives a translation factor of $-1.7380$ GHz / V for filter cavity of node B. **c** Transmission bandwidth characterization of the filter cavity of node A, obtained from Lorentzian fits to the transmission profiles of (**a**) after conversion to frequency using the translation factor from (a). Mean value of the cavity's transmission bandwidth was measured to be 66.08 MHz. **d** Corresponding data for node B filter cavity, using the translation factor from (**b**). The mean value of the cavity's transmission bandwidth for node B cavity was measured to be 75.25 MHz.

## SI3 Off-resonant excited photoluminescence spectra

The spectral properties of both emitters under off-resonant excitation were characterized using the calibrated filter cavities of SI2 as scanning spectrometers. Each $V_{Si}$ center was illuminated with continuous-wave off-resonant light at $728\,\mathrm{nm}$ while the voltage of its filter cavity in the zero-phonon line path was scanned, so that the transmitted single-photon counts trace the emission spectrum (Fig. SI3a, c). The reverse bias voltage of each node was set to the operating point of common emission frequency $\nu_{\mathrm{T}}$. The voltage axis was converted to relative detuning using the translation factors of SI2. The resulting double-peak transmission signal of the $A_1$ and $A_2$ transitions was fitted with a double Lorentzian. The measured profiles are the convolution of the emission lines with the cavity transmission. For two convolved Lorentzian profiles the full widths at half maximum add, so that the actual emission linewidth $\gamma_{\mathrm{emitter}}$ follows as

$$\gamma_{\mathrm{emitter}} = \gamma_{\mathrm{measured}} - \Gamma_{\mathrm{cavity}}. \qquad \text{(Eq. SI3)}$$

Here, $\gamma_{\mathrm{measured}}$ denotes the raw, measured Lorentzian fullwidth at half maximum, while $\Gamma_{\mathrm{cavity}}$ stands for the bandwidth of the used cavity. With measured $A_2$ linewidths of $95.7\,\mathrm{MHz}$ for $V_{Si,1A}$ and $202\,\mathrm{MHz}$ for $V_{Si,2B}$, and the individual cavity bandwidths $\Gamma_{\mathrm{A}} = 66.08\,\mathrm{MHz}$ and $\Gamma_{\mathrm{B}} = 75.25\,\mathrm{MHz}$ determined in SI2, this yields actual linewidths of $29.62\,\mathrm{MHz}$ for $V_{Si,1A}$ and $126.75\,\mathrm{MHz}$ for $V_{Si,2B}$, respectively.

To increase the fraction of emitted photons that can be collected, a $158\,\mathrm{nm}$ thick $SiO_2$ antireflective coating optimized for $916\,\mathrm{nm}$ was deposited on the sample surface (SI1). The saturation measurements were performed before and after the deposition. The same $V_{Si}$ centers were relocated after the coating step, so that the comparison refers to identical emitters rather than to a statistical ensemble. The dependence of the transmitted $A_2$ count rate on the off-resonant excitation power was measured for both emitters in both configurations (Fig. SI3b, d). The count rate follows a saturation behavior, with a saturation power $P_{\mathrm{sat}}$. Without the coating, the saturation powers are $P_{\mathrm{sat}} = 0.28\,\mathrm{mW}$ for $V_{Si,1A}$ and $0.40\,\mathrm{mW}$ for $V_{Si,2B}$. With the anti-reflective coating, the detected $A_2$ count rate at saturation increases by approximately a factor of 2 and the saturation power shifts to $0.51\,\mathrm{mW}$ and $0.74\,\mathrm{mW}$, respectively. All subsequent measurements were performed on the coated sample. Based on these curves, a common off-resonant excitation power of $1\ \mathrm{mW}$ was chosen for both nodes. This value deliberately lies above the saturation power of both emitters, which ensures maximum optical recycling of the color center. At the same time, the power was kept close to saturation rather than far above it, in order not to risk an increase of the optical linewidth at high off-resonant excitation. Operating on the plateau of the saturation curve additionally makes the emitted count rate insensitive to small drifts of the excitation power, which is essential over the $628\ \mathrm{h}$ campaign since the transmitted count rate serves as the error signal of the cavity locks (SI4.2).

At this excitation power of $1\ \mathrm{mW}$, the time-averaged transmitted $A_2$ count rate over the full interference campaign was $65\ \mathrm{cps}$ for each node, against a time-averaged background of $0.5\ \mathrm{cps}$ recorded at a laterally displaced position in the confocal scan under identical conditions (grey points in Fig. SI3a, c). This corresponds to signal-to-total count ratios of $S_1/I_1 = 0.99$ and $S_2/I_2 = 0.99$, which enter the cross-correlation model of SI6.

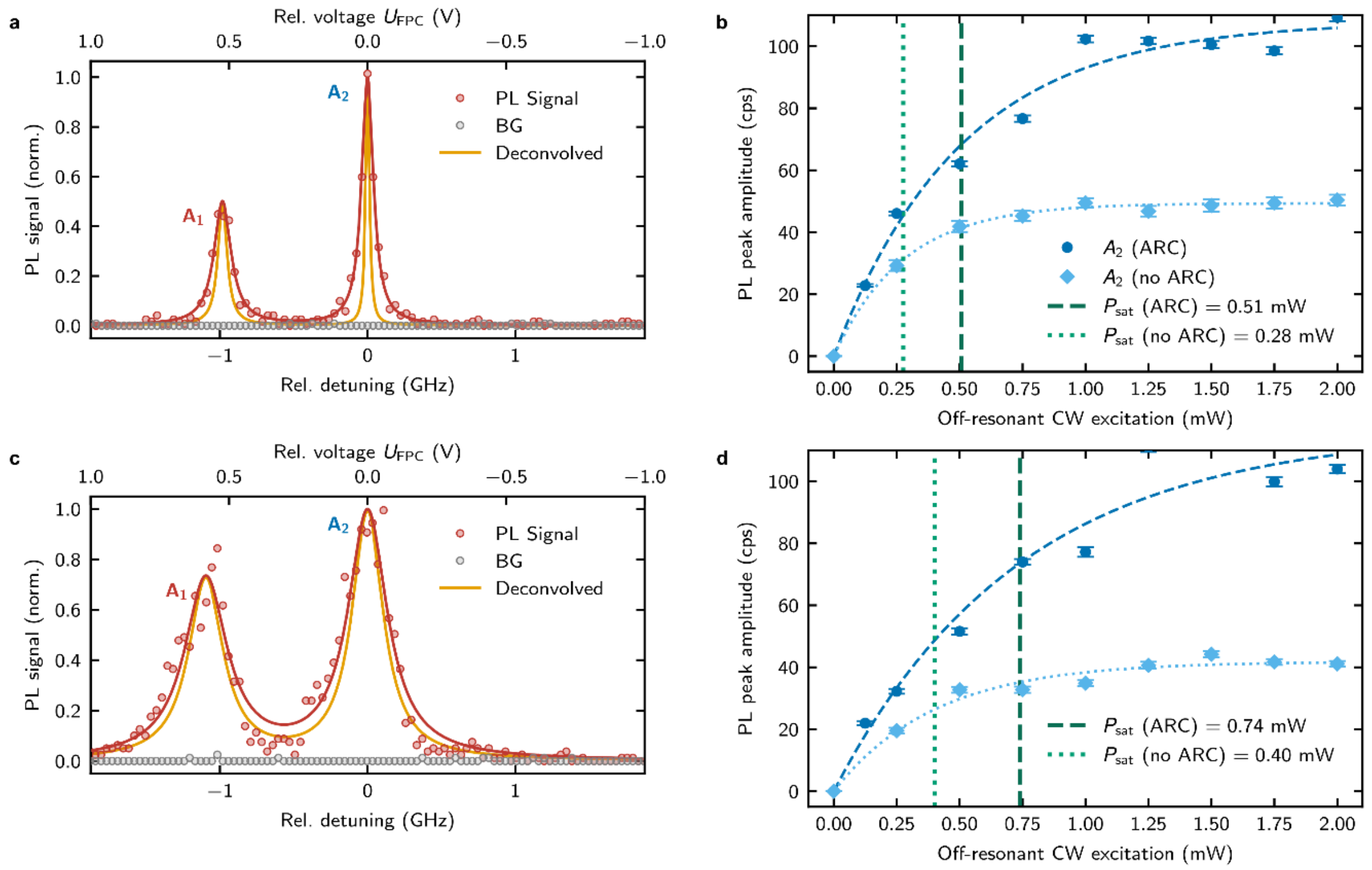

**Fig. SI3 Off-resonant photoluminescence spectra and saturation of $A_2$ emission of $V_{Si,1A}$ and $V_{Si,2B}$.**

**a** Cavity transmission spectrum of $V_{Si,1A}$ at an off-resonant excitation power of $1\ \mathrm{mW}$, recorded at the reverse bias voltage of common $A_2$ emission frequency $\nu_T$. Red points denote normalized single-photon transmission counts, grey points the background measured at a laterally displaced position in the confocal scan at same power. The red line is a double-Lorentzian fit to the data, the yellow line the deconvolved fit according to Eq. SI3. The upper axis denotes the applied cavity voltage. **b** $A_2$ emission saturation curve of $V_{Si,1A}$ at different off-resonant excitation powers, measured before (light blue diamonds) and after (dark blue circles) deposition of the $SiO_2$ antireflective coating. Dashed vertical lines mark the respective saturation powers before (light green) and after (dark green) $SiO_2$ deposition. **c, d** Corresponding data for $V_{Si,2B}$.

## SI4 Automated operation and stabilization of experimental setup

### SI4.1 Automated long-term operation

Maintaining two-photon interference over $628\,\mathrm{h}$ requires two conditions to hold simultaneously: the resonance of each Fabry-Pérot filter cavity must stay centered on the $A_2$ emission of its node, and the $A_2$ transition frequencies of the two nodes must stay equal at the common target frequency $\nu_\mathrm{T}$. Both drift independently. The cavity resonance drifts through thermal expansion and relaxation of the piezoelectric actuator and the emission frequency through changes in the local charge environment of the particular color center. Manual intervention would disturb the confocal alignment and the cavity locks, so the entire sequence was executed by a control script without user interaction.

Both drifts appear in the same observable: a decrease of the transmitted single-photon count rate. The count rate alone therefore cannot reveal whether the cavity has moved away from the emitter or the emitter away from the target frequency $\nu_\mathrm{T}$ and thus as well as from the cavity. The experiment consequently employs two control loops that differ both in timescale and in the quantity they measure. The fast loop, executed every $10\,\mathrm{s}$, keeps each cavity centred on the maximum of its transmission by regulating the piezo voltage, using the transmitted count rate as the error signal (SI4.2). It corrects cavity drift, but it is blind to the absolute optical frequency. Losing its stable operation and thus the transmitted $A_2$ counts is referred further in this section as a cavity lock loss. The slow loop, executed every $2\,\mathrm{h}$, records a resonant PLE spectrum at each node and thereby determines the absolute $A_2$ emission frequency. Any deviation from $\nu_\mathrm{T}$ is corrected through the reverse bias voltage of the *p*-*i*-*n* diode (SI7). This is the only step in the sequence that yields an absolute frequency rather than a count rate. The remainder of this section describes how both loops are embedded into the automated measurement sequence. The corresponding control architecture is shown in Fig. SI4.1. The total measurement campaign of $628\,\mathrm{h}$ is composed of $29$ measurement runs, each of which is subdivided into individual sessions. A session is the unit over which correlation data is accumulated without interruption. It is terminated either when a cavity lock is lost, in which case a full realignment follows, or when the timer of the slow frequency loop expires and the $A_2$ emission frequency of both nodes is verified (SI7).

This timer counts active correlation time rather than wall-clock time, so that the $2\,\mathrm{h}$ interval refers to accumulated measurement and is not consumed by realignment, locking or frequency calibration. The correlation histograms of all runs are summed in post-processing, which allows the total measurement time to be built up cumulatively from individually verifiable runs. A run log is written separately for each session, so that a session can be inspected, stopped and restarted without affecting the others.

A measurement run begins with hardware initialization. All four channels of the source-measure unit (SMU) are set to their starting values: CH1 and CH4 supply the reverse bias voltages of the *p-i-n* samples at nodes A and B, while CH2 and CH3 drive the piezoelectric actuators of the Fabry-Pérot filter cavities individually. Both off-resonant $728\,\mathrm{nm}$ lasers are initialized in standby. Two time-tagged channels DCH1 and DCH2 are opened and ready for measurement. They are connected to the detectors at the two beam splitter outputs.

The subsequent alignment and initial-lock sequences are executed for both setups subsequently. First, a three-dimensional confocal optimization maximizes the phonon-sideband count rate of each emitter, compensating any positional lateral or axial drift. Then, for each setup in turn, the off-resonant laser is switched to continuous-wave operation (CW) and the cavity piezo-voltage is swept across the emission signal of the observed $V_{Si}$ while the transmitted counts are recorded in sum on DCH1 and DCH2. The maximum of the scanned peak, if it exceeds a preset minimum threshold, defines the setpoint of the fast loop for that cavity. Once engaged, the loop holds the cavity at this setpoint for the remainder of the session, so that the filter cannot drift off the $A_2$ line while correlation data is being accumulated. Setup A is locked first, followed by setup B. During the peak scan of setup B, the laser of setup A is switched off, so that the detected counts originate from node B alone. The main measurement loop is multithreaded and runs three concurrent tasks: the interleaved PID lock of the two cavities (SI4.2), acquisition and periodic saving of the correlation data together with live plotting, and a continuous health check of all hardware connections and threads. With both lasers in CW operation and both cavity setpoints held in a preset window of possible PID-error, the cross-correlation $G_{12}^{(2)}(\tau)$ between DCH1 and DCH2 is accumulated by the time-tagging electronics. Two decision nodes are evaluated continuously. The first monitors the summed count rate of both detector channels against its total setpoint of approximately $130\,\mathrm{cps}$, corresponding to the summed setpoint of the $65\ \mathrm{cps}$ contributed by each node. A deviation exceeding $\pm 50\ \mathrm{cps}$ from the total setpoint triggers the lock-loss handler. This criterion is two-sided, as a decrease indicates that one of the cavities has lost transmission, while an increase indicates stray light entering the setup, for example when the enclosure is opened. This is a failure criterion rather than the regulation tolerance of the fast loop, which corrects far smaller deviations continuously (SI4.2). The second decision node verifies that both PID loops are still active and that all hardware components respond. If either condition fails, control passes to the lock-loss handler, which saves the current session data and triggers a full realignment, i.e. renewed confocal optimization for both setups followed by re-determination of both PID setpoints.

Independently of the lock status, a timer triggers the periodic frequency calibration after every 2h of accumulated correlation time. The sequence measures a resonant PLE spectrum at each node, extracts the $A_2$ transition frequency and, if required, adjusts the reverse bias voltage via CH1 for setup A or CH4 for setup B. The procedure is described in detail in SI7. Once, both nodes lie within the acceptance window, a full realignment is triggered, subsequently the correlation measurement resumes and the timer restarts. The implementation of the fast stabilization loop, which keeps each cavity on the transmission of its own node, is described in SI4.2. Correspondingly, the slow stabilization loop, which keeps the two emitters on the common target emission frequency $\nu_{\mathrm{T}}$, is described in SI7. Together they cover the two independent drifts identified above.

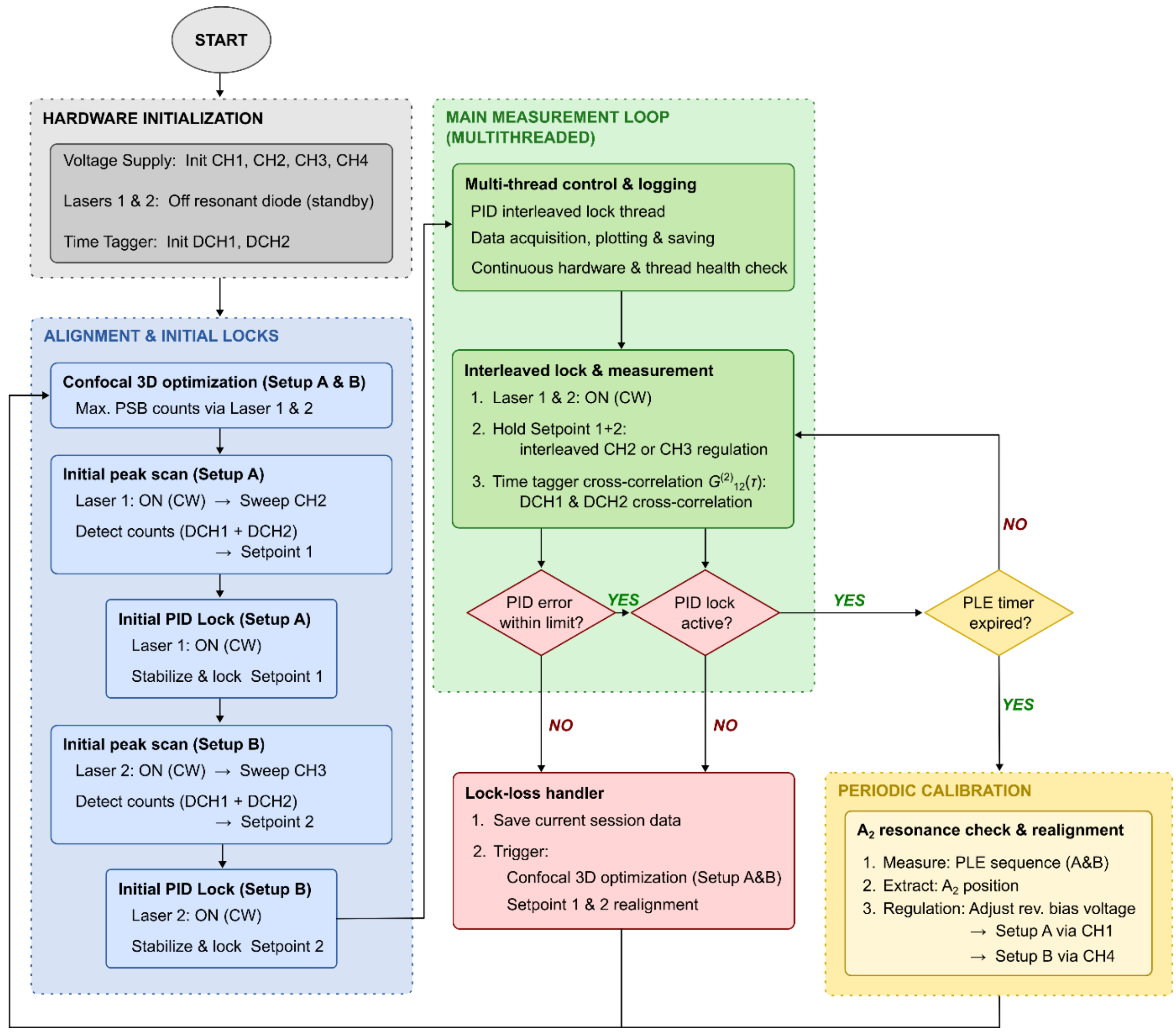

**Fig. SI4.1 Operation flowchart and control architecture for automated two-photon-interference experiment.**

Measurement session starts with hardware initialization (voltage channels CH1–CH4, off-resonant lasers, and time tagger channels DCH1/DCH2). Voltage channels CH1 and CH4 are used for the reverse bias voltage of the nodes' A and B *p-i-n* samples. CH2 and CH3 are used for the piezo-electric Fabry-Pérot cavity control, individually. The alignment & initial locks sequence (blue) establishes spatial confocal 3D optimization and establishes PID lock setpoints (Setpoint 1 & 2) sequentially for Setup A and B. The main multithreaded measurement loop (green) concurrently handles data acquisition, plotting, hardware health monitoring, interleaved PID feedback stabilization (via CH2 and CH3), and second-order photon correlation $G^{(2)}_{12}(\tau)$ acquisition. Active decision nodes (red) continuously verify PID lock status and error limits. A lock loss saves the current session and re-initiates full alignment. Upon expiration of the periodic PLE timer, the sequence executes an $A_2$ resonance calibration (yellow) to adjust reverse bias voltages (CH1 for Setup A, CH4 for Setup B) before resuming measurement. Both the lock-loss handler and the periodic calibration return to the confocal 3D optimization, so that each of the two termination conditions starts a new session. Boxes denote process blocks; diamonds denote decision nodes. CW, continuous wave regime; PID, proportional-integral-derivative control; PLE photoluminescence excitation emission spectra; PSB phonon sideband emission.

### SI4.2 Fast stabilization loop: interleaved PID regulation of Fabry-Pérot cavities

Each filter cavity must remain centered on the $A_2$ transition of its node for the entire duration of a session. With transmission bandwidths of $66.08\,\mathrm{MHz}$ for node A and $75.25\,\mathrm{MHz}$ for node B (SI2), the cavities are narrow compared with the drift of their piezoelectric actuators, so that even a small uncorrected change of the piezo voltage moves the cavity off resonance and reduces the transmitted count rate. This is apparent already during alignment: when a cavity is set to the transmission maximum of the stabilized narrow-band reference laser and the piezo voltage is subsequently held, the transmitted count rate begins to decrease within seconds. Both cavities were therefore regulated continuously throughout the campaign.

The only signal available for this regulation during the interference measurement is the transmitted photon rate itself, registered as the sum of the two time-tagger channels DCH1 and DCH2. This creates a difficulty. Photons from both nodes are combined on the beam splitter before detection, so each detector receives light from both cavities and the summed count rate does not reveal which of the two has drifted. Operating both loops at the same time would mean that each loop responds to changes caused by the other. The two lock loops are therefore separated in time rather than in signal. At any moment only one cavity is actively regulated, while the piezo voltage of the other is held at its last value. Any change of the summed count rate during an active interval can then be attributed unambiguously to the cavity currently under regulation. Since the held cavity remains close to resonance over such a short time interval, both nodes continue to transmit photons throughout, and the accumulation of the cross-correlation $G_{12}^{(2)}(\tau)$ is never interrupted by the regulation. The scheme is shown in Fig. SI4.2. Regulation alternates between the two nodes in intervals of $10\,\mathrm{s}$. During an active interval, the deviation of the measured count rate from the setpoint forms the error signal $e_i$, which is processed by the PID controller of node $i$ with proportional gain $K_\mathrm{P}$, integral gain $K_\mathrm{I}$ and derivative gain $K_\mathrm{D}$, updated at a rate of $0.5\ \mathrm{Hz}$ in order to suppress short-term shot noise. Its output $c_i$ is converted into the piezo voltage $u_i$, applied through CH2 for the cavity of setup A and CH3 for that of setup B. At the end of the interval the controller output is frozen, the last applied voltage is held, and the loop of the other node is activated. The setpoints are determined once during the initial locking sequence and are not modified during a session (SI4.1). Rather than the transmission maximum itself, a point at $95\ \%$ of the maximum of the cavity transmission scan on the $A_2$ peak is chosen. At the maximum the derivative of the transmission with respect to the piezo voltage vanishes, so that a deviation in both directions reduces the count rate identically. Thus, the error signal carries no directional information at the transmitted peak's maximum. On the flank the transmission varies monotonically with the piezo voltage, so that the sign of the deviation from the setpoint identifies the direction in which the cavity has drifted. This is what allows the count rate alone to serve as an error signal for the loop. The measurement is interrupted only if the summed count rate of both nodes (Setpoint A+B) deviates from its setpoint by more than $\pm 50\ \mathrm{cps}$, which may indicate that one of the cavities has lost transmission entirely. In this case the control script triggers a full realignment (SI4.1).

The residual performance of the interleaved lock was quantified from the distributions of the PID control voltage residuals recorded over the whole campaign, comprising approximately $400{,}000$ samples for each node. Converted to frequency using the cavity calibration of SI2, these correspond to root-mean-square residual detuning of $4.29\ \mathrm{MHz}$ for node A and $3.93\ \mathrm{MHz}$ for node B, i.e. below 10 % of the respective filter bandwidth. The resulting reduction of the interference contrast is evaluated in SI6. The gains $K_\mathrm{P}$, $K_\mathrm{I}$ and $K_\mathrm{D}$ were adjusted following the standard procedure for PID initialization[2], in which the proportional gain is set first, followed by the integral and finally the derivative gain.

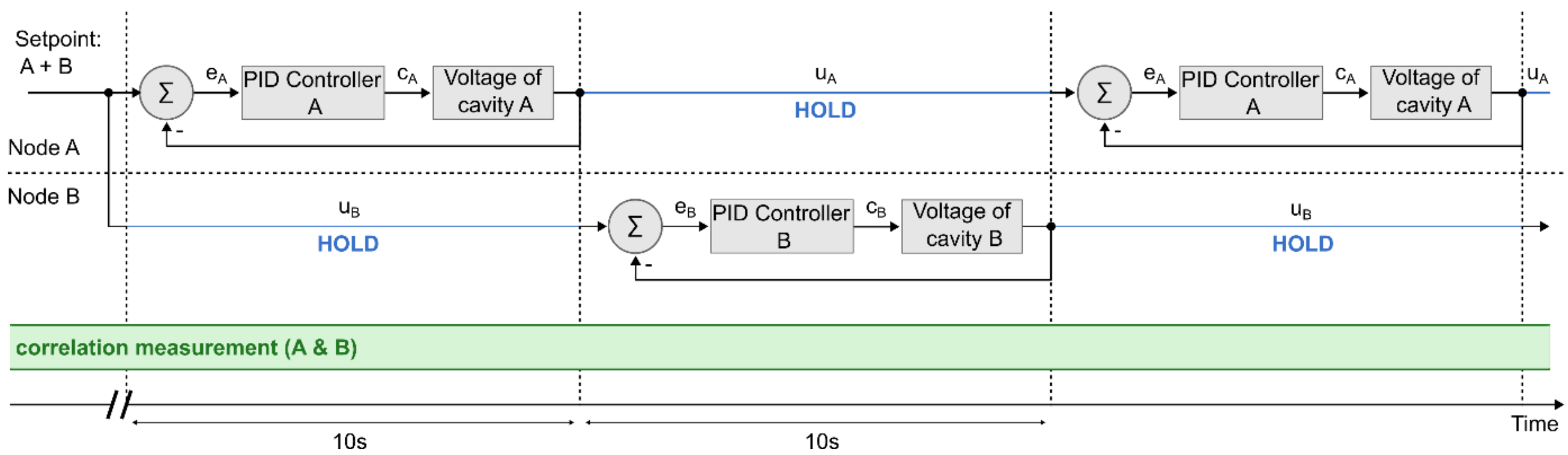

**Fig. SI4.2 Principle of the interleaved PID regulation of Fabry-Pérot cavity voltages.**

Control scheme over time for nodes A and B. The regulation alternates between node A (top) and node B (bottom) in intervals of $10\ \mathrm{s}$. In an active interval, the deviation of the transmitted count rate from the setpoint (A+B) forms the error signal $e_i$, which the PID controller converts into the control output $c_i$ and hence the piezo voltage $u_i$ of the respective cavity. During the inactive interval the voltage of the other node is held at its last value (bluish HOLD), so that any change of the detected count rate can be attributed unambiguously to the cavity currently under regulation. The accumulation of the cross-correlation between both nodes, marked in green, proceeds continuously throughout. Black arrows denote active operation, blue arrows show passively held operation.

## SI5. Single photon emission characteristics

### SI5.1 Second-order autocorrelation of the phonon sideband emission

The single-photon character of both emitters $V_{Si,1A}$ and $V_{Si,2B}$ was verified in a Hanbury Brown–Twiss (HBT) configuration prior to the interference campaign. The measurements were carried out on the phonon sideband (PSB) rather than on the zero-phonon line, because the PSB carries the larger fraction of the total emission. A commercially available $50{:}50$ single mode 1x2 fiber-coupler (Thorlabs, custom made for 916nm) was used therefore for a standard HBT arrangement, in which the PSB emission of a single node is split onto the two superconducting nanowire single-photon detectors. Detection events were time-tagged and plotted in a time-difference histogram between counting events with a bin width of $200\ \mathrm{ps}$ on a total measurement window of $\pm 400\ \mathrm{ns}$ (Fig. SI5.1). Both nodes were excited off-resonantly at 728 nm with an optical power of $1\ \mathrm{mW}$, under identical conditions of the subsequent two-photon interference measurement, so that the resulting values quantify the single-photon purity in the regime actually used for the HOM experiment. Based on the framework of ref.[3], to remove influence of background noise, the measured autocorrelation data $g^{(2)}_{ii,\mathrm{PSB}}(\tau)$ of each node $i$ is was corrected to background using $g^{(2)}_{ii}(\tau) = (g^{(2)}_{ii,\mathrm{PSB}}(\tau) - (1-\rho^2))\,/\rho^2$, where denotes $\rho = S/(S+B)$ with $S$ as the total signal counts and $B$ as background counts. Using the expression $g^{(2)}_{ii}(\tau) = 1 - (1+a_i)\exp\left(-\frac{|\tau|}{\tau_{1,i}}\right) + a_i \exp\left(-|\tau|/\tau_{2,i}\right)$ for the ideal autocorrelation function, the data was fit to the following formula:

$$g^{(2)}_{ii,\mathrm{PSB}}(\tau) = 1 - \rho_i^2 \left\{(1+a_i)\mathrm{e}^{\left(-\frac{|\tau|}{\tau_{1i}}\right)} - a_i \mathrm{e}^{\left(-\frac{|\tau|}{\tau_{2i}}\right)}\right\} \qquad \text{(SI5.1)}$$

Here, $\tau_1$ denotes the antibunching time constant governed by the excitation rate and the excited-state lifetime, $\tau_2$ the bunching time constant associated with shelving into the metastable state, and $a$ the bunching amplitude.

The parameter $\rho_i$ was measured independently for each observer $V_{Si}$ on a time-averaged CW measurement. For $V_{Si,1A}$ this factor was $\rho_1 = 0.95$ and for $V_{Si,2B}$ $\rho_2 = 0.93$. For $V_{Si,1A}$ this yields $g^{(2)}_{11,\mathrm{PSB}}(0) = 0.0930(5)$ (Fig. 2b), and for $V_{Si,2B}$ $g^{(2)}_{22,\mathrm{PSB}}(0) = 0.1293(5)$ (Fig. SI5.1). Both values lie well below the classical two-photon threshold of $0.5$, confirming that each node emits single photons.

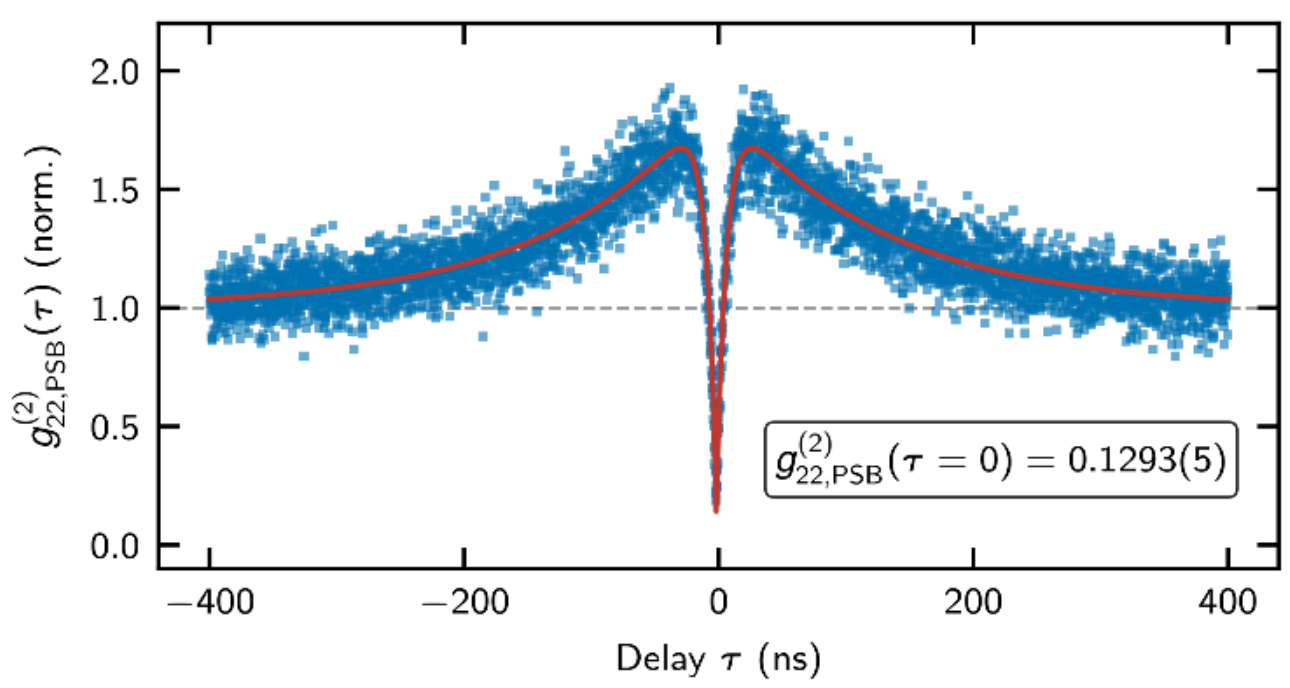

**Fig. SI5.1 Second-order autocorrelation measurement in Hanbury Brown-Twiss configuration of $V_{Si,2B}$.**

Background-corrected autocorrelation measurement of the phonon sideband (PSB) emission of $V_{Si,2B}$ in Hanbury Brown-Twiss configuration under off-resonant excitation at $728\,\mathrm{nm}$ and $1\mathrm{mW}$. Fit of normalized data yields $g^{(2)}_{22,\mathrm{PSB}}(0) = 0.1293(5)$. Blue squares are the background-corrected data; red solid lines represent the fit of the three-level rate-equation model of Eq. SI5.1. Value at zero delay derived from the fit. Corresponding measurement for $V_{Si,1A}$ in main text Fig. 2b.

## SI5.2 Resonant optical linewidths and lifetime limit

The optical linewidths quoted in the main text refer to the reverse bias voltages at which both emitters share the common target emission frequency $\nu_{\mathrm{T}} = 327.231000\ THz$, i.e. to the operating point of the interference experiment. At this operating point, resonant photoluminescence excitation yields Lorentzian full widths at half maximum for the $A_2$ emission of $\gamma_{1\mathrm{A}} = 34.26 \pm 0.92\ \mathrm{MHz}$ for $V_{Si,1A}$ and $\gamma_{2\mathrm{B}} = 58.83 \pm 3.11\ \mathrm{MHz}$ for $V_{Si,2B}$ (Fig. SI5.2a,b). Each spectrum was recorded in a dual-tone scheme with a carrier-excitation power of $5\ \mathrm{nW}$. While sweeping the applied reverse bias voltage of the *p-i-n* diode, a reduction of the linewidths is observed. This effect refers to the reduction of free charges, due to an increasing local electric field in the vicinity of the observed $V_{Si}$ centers. For $V_{Si,1A}$ the linewidth narrows from $96.54 \pm 5.20\ \mathrm{MHz}$ at zero bias to $32.31 \pm 1.11\ \mathrm{MHz}$ at $30\ \mathrm{V}$ reverse bias. For the operation point of common emission frequency $\nu_{\mathrm{T}}$ a reverse bias voltage of $18.632\ \mathrm{V}$ was applied. Correspondingly for $V_{Si,2B}$, the linewidth narrows down from $134.68 \pm 10.63\mathrm{MHz}$ at zero bias to $46.91 \pm 2.34\ \mathrm{MHz}$. Reverse bias voltage for common $A_2$ emission frequency was set to $18.623\ \mathrm{V}$. To determine the Fourier-limit of these emitters, the excited-state lifetime of each emitter was measured independently. Each $V_{Si}$ was excited resonantly at $A_2$ resonance with $< 1\ \mathrm{ns}$ pulses, generated by the combination of an acousto-optic modulator and an amplitude electro-optic modulator driven by a programmable pulse generator. The arrival times of the emitted photons relative to the excitation pulse were measured (Fig. SI5.2c) accordingly. Linear fits to the logarithmically normalized decay yield $\tau_{1,1\mathrm{A}} = 11.06 \pm 0.09\ \mathrm{ns}$ for $V_{Si,1A}$ and $\tau_{1,2\mathrm{B}} = 11.17 \pm 0.20\ \mathrm{ns}$ for $V_{Si,2B}$. These values are consistent with the $A_2$ excited-state lifetime of $11.3\ \mathrm{ns}$ reported for V2 centers in bulk 4H-SiC[4,5].

A radiative lifetime $\tau$ sets a lower bound on the optical linewidth of $\gamma = 1/(2\pi\tau)$. The measured lifetimes therefore correspond to Fourier limits of $\gamma_{1\mathrm{A,FL}} = 14.39 \pm 0.12\ \mathrm{MHz}$ and $\gamma_{2\mathrm{B,FL}} = 14.25 \pm 0.26\ \mathrm{MHz}$. Comparing these with the resonant linewidths at the operating point, $V_{Si,1A}$ exceeds its lifetime limit by a factor of $2.4$ and $V_{Si,2B}$ by a factor of $4.1$.

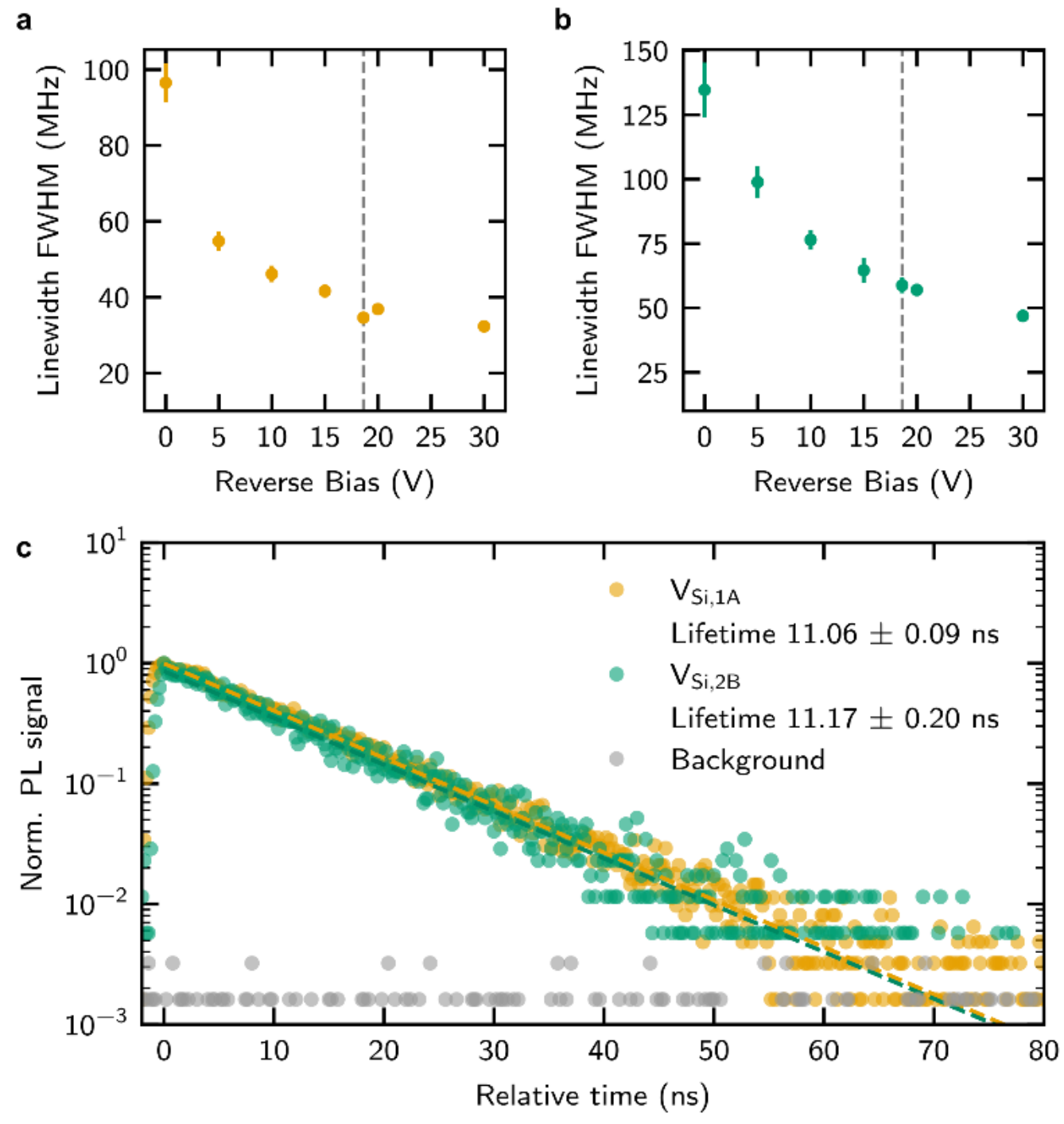

**Fig. SI5.2 Linewidth reduction and excited-state lifetimes of $V_{Si,1A}$ and $V_{Si,2B}$ on the $A_2$ transition.**

**a** Optical linewidth (FWHM) of $V_{Si,1A}$ as a function of applied reverse bias voltage, extracted from Lorentzian fits to dual-tone resonant photoluminescence excitation spectra of the $A_2$ transition recorded at a carrier excitation power of $5\ \mathrm{nW}$. Reverse bias operation point for common emission frequency is marked with a grey dashed vertical line. **b** Corresponding measurement for $V_{Si,2B}$. Error bars in (**a**) and (**b**) denote the standard error of Lorentzian fit. **c** Excited-state lifetime of $V_{Si,1A}$ and $V_{Si,2B}$ at operation point for common target emission frequency, under pulsed resonant excitation of the $A_2$ transition with pulses shorter than $1\ \mathrm{ns}$. Photon arrival times relative to the excitation pulse are shown on logarithmic intensity scale for $V_{Si,1A}$ (yellow) and $V_{Si,2B}$ (green), together with detuned measured background signal (grey).

## SI6 Cross-correlation function of two-photon interference

The measured cross-correlation data is modelled and fit with the following function of the two-photon interference:

$$G_{12}^{2(\tau)} = c_1^2 g_{11}^2(\tau) + c_2^2 g_{22}^2(\tau) + 2c_1 c_2 \cdot \{1 - \eta \cdot \frac{S_1 S_2}{I_1 I_2} \cdot |g_{11}^{(1)}(\tau)| \cdot |g_{22}^{(1)}(\tau)| \cdot \cos(\Delta\omega_0 \tau) \cdot \exp\left(-\frac{\Gamma \cdot \tau}{2}\right) \exp\left(-\frac{\sigma^2 \cdot \tau^2}{2}\right)\} \quad \text{(SI6.1)}$$

This function is based on previous performed experiments of this kind[6–10]. The first two summand terms describe the autocorrelation contribution of each emitter, weighted by the relative count rates $c_i = I_i / (I_1 + I_2)$, where $I_i$ denotes the total emitter's $i$ emission intensity, thus pure signal $S_i$ counts and background counts $B_i$ together as $I_i = S_i + B_i$. The autocorrelation terms are modeled as ideal terms $g_{ii}^{(2)}(\tau) = 1 - (1 + a_i) \exp\left(-\frac{|\tau|}{\tau_{1,i}}\right) + a_i \exp(-|\tau|/\tau_2)$, with $\tau_{1,i}$ being the coherence time of the emitter $i$ governed by the excited-state lifetime, $\tau_2$ the common bunching time constant and $a$ the common bunching amplitude. The third summand implies the interference. Within it, the spectral properties of the emitters enter through their first-order coherence functions (field autocorrelation) $\left|g_{ii}^{(1)}(\tau)\right| = \exp(-|\tau|/2\tau_{1,i})$, the mutual, mean detuning through $\cos(\Delta\omega_0 \tau)$ with $\Delta\omega_0 = \omega_1 - \omega_2$, and the session-to-session long term spectral diffusion through a Gaussian kernel as $\exp\left(-\frac{\sigma^2 \tau^2}{2}\right)$ with its scattering pregiven by the parameter $\sigma$. The finite spectral bandwidth transmitted by the Fabry-Pérot filter stage is accounted for by the Lorentzian term $\exp(-\Gamma\tau/2)$, with $\Gamma$ given as an angular frequency. The two cavities differ slightly in their measured transmission bandwidths, $66.08\ \mathrm{MHz}$ for node A and $75.25\ \mathrm{MHz}$ for node B (SI2). Since a single common value enters Eq. SI6.1, $\Gamma$ was not treated as a free fit parameter but fixed to the larger of the two, $\Gamma = 2\pi \cdot 75.25\ \mathrm{MHz}$. This maximizes the damping term and therefore yields the narrowest coherence envelope compatible with the measured filter characteristics, so that the resulting interference contrast is a conservative estimate: any smaller bandwidth would widen the envelope and increase the modelled contrast. With this choice, the fit of Eq. SI6.1 to the measured cross-correlation yields a reduced chi-square of $1.465$ and a cross-correlation minimum of $G_{12,\parallel}^{(2)}(0) = 0.0914 \pm 0.0030$ for the indistinguishable cross-correlation measurement.

The phenomenological prefactor $\eta$ collects all remaining technical imperfections of the interferometric setup i.e., every effect that reduces the interference contrast but is not already contained in the spectral properties of the two emitters. This parameter was not treated as a free fit parameter. Each contribution was determined independently and $\eta$ is their product,

$$\eta = \eta_{\mathrm{BS}} \cdot \eta_{\mathrm{POL}} \cdot \eta_{\mathrm{SMO}} \cdot \eta_{\mathrm{CAV}}. \quad \text{(Eq. SI6.2)}$$

Taking the contributions of beam splitter imbalance (BS), polarization mismatch (POL), spatial mode overlap (SMO) and residual detuning of filter cavities (CAV) into account, the total product of $\eta$ yields $\eta = 0.9991 \text{ x } 0.9773 \text{ x } 0.9968 \text{ x } 0.9891 = 0.9627 \ \cong 0.963$. The detailed calculation of these factors is explained further in the following paragraphs.

**Beam splitter imbalance $\boldsymbol{\eta}_{\mathbf{BS}}$.**

A splitting ratio deviating from unity reduces the interference contrast, because both interfering paths do not contribute with equal amplitude in such a case. Following the correction procedure established for two-photon interference experiments[11], the contrast reduction can be expressed by

$$\eta_{\mathrm{BS}} = \frac{2RT}{(R^2+T^2)} = \frac{2R(1-R)}{1-2R+2R^2}, \quad \text{(Eq. SI6.3)}$$

where $R$ is the reflectivity and $T = 1 - R$ the transmissivity of the beam splitter. The splitting ratio of the free-space beam splitter was determined from the collected counts on each detector channel of the cross-correlation measurement itself, yielding time-averaged values of $\langle R_{\parallel} \rangle = 0.5108$ for the co-polarized configuration (indistinguishable) in which the photons interfere, and $\langle R_{\perp} \rangle = 0.5136$ for the orthogonally polarized reference configuration. Since $\eta$ enters Eq. SI6.1 only for the indistinguishable case, the co-polarized value applies and gives $\eta_{\mathrm{BS}} = 0.9991$.

**Polarization mismatch $\boldsymbol{\eta}_{\mathbf{POL}}$.**

Two photons in different polarization states are partially distinguishable. For two single photons described by density matrices $\boldsymbol{\rho}_1$ and $\boldsymbol{\rho}_2$, the Hong-Ou-Mandel interference visibility is given by their overlap $V = \mathrm{Tr}\{\boldsymbol{\rho}_1 \boldsymbol{\rho}_2\}$[12], a relation that holds for a two-level system and is therefore directly applicable to the polarization degree of freedom. Writing each polarization state in the Bloch representation $\boldsymbol{\rho}_i = 1/2 \ \cdot (\mathbb{1} + \mathbf{s}_{\mathrm{i}} \cdot \boldsymbol{\sigma})$ with Stokes vectors $\boldsymbol{s}_i$ and the vector of Pauli matrices $\boldsymbol{\sigma}$, this evaluates to

$$\eta_{\mathrm{POL}} = \mathrm{Tr}\{\rho_1 \rho_2\} = \frac{1}{2}(1 + \boldsymbol{s}_1 \cdot \boldsymbol{s}_2). \quad \text{(Eq. SI6.4)}$$

The Stokes vectors are normalized such that $|\boldsymbol{s}_i|$ equals the degree of polarization of node $i$, so that Eq. SI6.4 reduces to $1/2$ for a fully unpolarized state and to unity for identical, fully polarized states. Both arms in the interferometric setup were characterized with a polarimeter (Thorlabs PAX1000IR2), using the resonant laser at common emission frequency $\nu_\mathrm{T}$ as a reference. Arm 1 gave an azimuth of $0.02°$, ellipticity of $0.13°$ and a degree of polarization of $98.18\%$ at $10\ \mu\mathrm{W}$ illumination power. Arm 2 had an azimuth of $-0.01°$, an ellipticity of $0.11°$ and a degree of polarization of $97.23\%$ at same illumination power. These values correspond to $\boldsymbol{s}_1 = (0.981790, 0.000685, 0.004455)$ and $\boldsymbol{s}_2 = (0.972293, -0.000339, 0.003733)$. This yields in total a reduction of interference contrast induced by polarization mismatch of $\eta_\mathrm{POL} = 0.9773$.

**Spatial mode overlap $\boldsymbol{\eta_{SMO}}$.**

Since the two photons are overlapped on a free-space beam splitter, an imperfect match of their mode profiles reduces the interference contrast. The intensity profiles of both beams were recorded and fitted with two-dimensional Gaussian profiles (Fig. SI6a,b), giving a spatial mode overlap of $\eta_\mathrm{SMO} = 0.9968$.

**Residual detuning of filter cavities $\boldsymbol{\eta_\mathrm{CAV}}$.**

Both Fabry-Pérot filter cavities are stabilized during operation (SI4.2), but the residual jitter of the PID controlled voltage translates into a fluctuating detuning between the two transmitted spectra. For two exponentially decaying wave packets with Lorentzian spectra of full-width half maxima $\Gamma_{i,j}$ and a mutual detuning $\delta\nu$, the spectral overlap evaluates to

$$\eta_\mathrm{CAV} = \frac{4\Gamma_i\Gamma_j}{\left(\Gamma_i+\Gamma_j\right)^2+4\delta\nu^2} \qquad \text{(Eq. SI6.5).}$$

With the measured filter bandwidths $\Gamma_A = 66.08\ \mathrm{MHz}$ and $\Gamma_\mathrm{B} = 75.25\ \mathrm{MHz}$ (SI2) and the root-mean-square residual detuning between the two cavities, $\delta\nu = \sqrt{(\delta\nu_1{}^2 + \delta\nu_2{}^2)} = \surd(4.29^2 + 3.93^2)\ \mathrm{MHz} = 5.82\ \mathrm{MHz}$, Eq. SI6.5 gives $\eta_\mathrm{CAV} = 0.9891$. Note that the two cavities contribute a small reduction of $0.42\ \%$ even at zero detuning, since their bandwidths are not identical. The individual values $\delta\nu_1 = 4.29\ \mathrm{MHz}$ and $\delta\nu_2 = 3.93\ \mathrm{MHz}$ were obtained from the voltage jitter distributions of the PID control voltage residuals over all interleaved PID sessions (Fig SI6c,d) for the indistinguishable measurement run, converted to frequency using the cavity calibration of SI2.

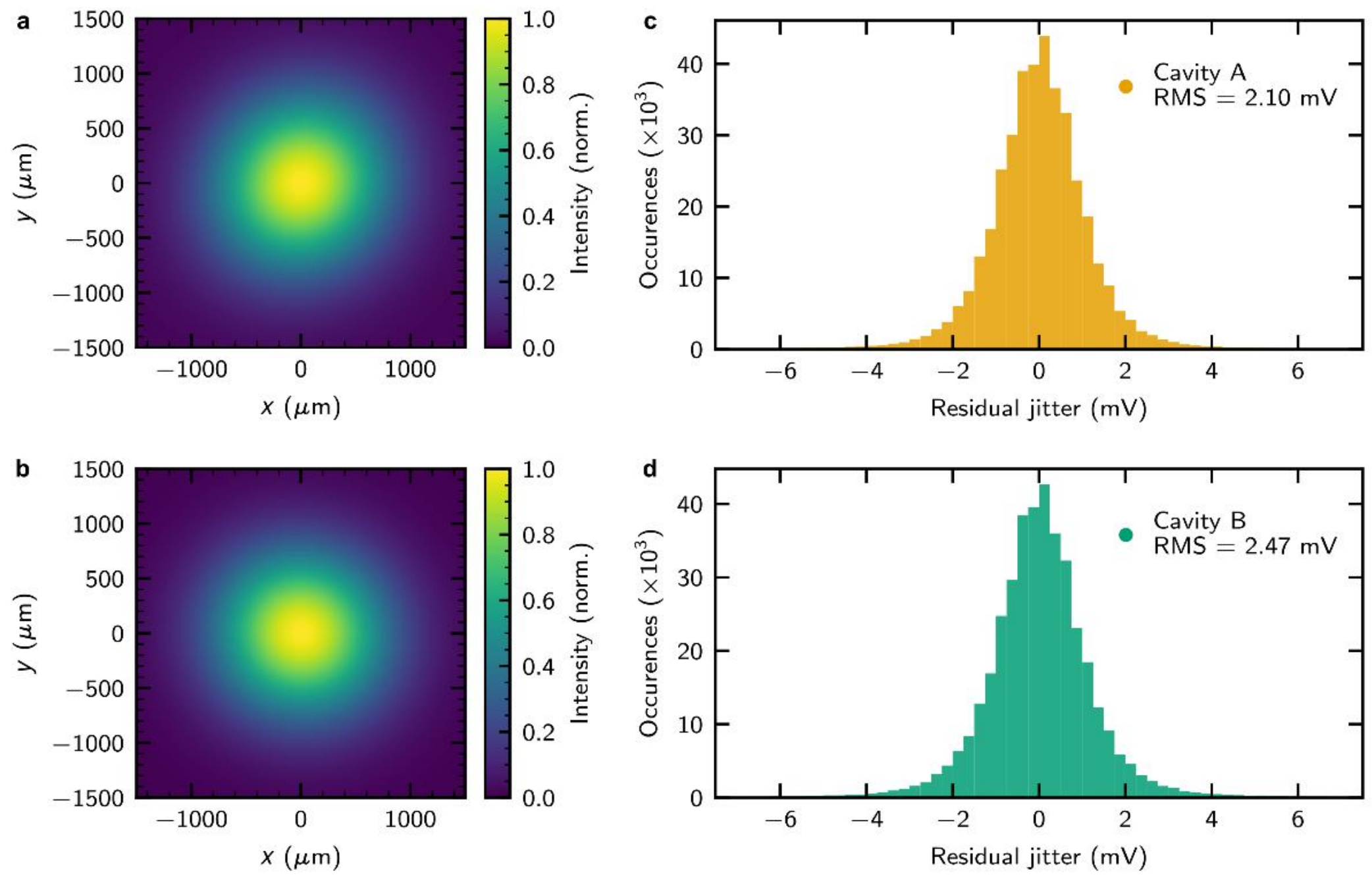

**Fig SI6 Interferometric setup impurities: spatial modes and fast loop PID voltage jitter residues of zero-phonon line path of $V_{Si,1A}$ and $V_{Si,2B}$.**

**a** Normalized intensity beam profile of zero-phonon line path of node A. Beam profile recorder at a position equivalent to that of the $50:50$ beam splitter, where both signal paths interfere. Figure shows corresponding Gaussian fit of recorded intensity profile. **b** Corresponding measurement of beam profile of ZPL path of node B. **c** Distribution of residual voltage jitter of fast loop PID control for node A cavity. Voltage jitter distribution accumulated from the $10\,s$ control windows over the whole measurement campaign of $628\,\mathrm{h}$ for the cross-correlation of indistinguishable photons. The root-mean-square (RMS) value of this distribution is $2.10\,\mathrm{mV}$, corresponding to a residual detuning of $3.93\,\mathrm{MHz}$ at the voltage-to-frequency translation factor of $-1.8745\,\mathrm{GHz/V}$ (SI2) **d** Corresponding distribution for the cavity of node B. RMS of voltage jitter is here $2.47\,\mathrm{mV}$ and equals to a residual detuning of $4.29\,\mathrm{MHz}$ at $-1.7380\,\mathrm{GHz/V}$.

## SI7 Slow stabilization loop: remote control of the emission frequency

To keep the $A_2$ transition of both emitters at the common target frequency $\nu_T$ and to compensate its slow drift, the emission frequency of each node was verified periodically every $2\,\mathrm{h}$ of active accumulated correlation time (SI4.1). Without correction, the emission frequency of either $V_{Si}$ center was found to drift by up to $\pm 40$ $\mathrm{MHz}\,/\mathrm{h}$ (on a timescale of 2 hours), which sets the required checking interval. Each verification consists of a resonant PLE scan of the $A_2$ transition over a range of $3\,\mathrm{GHz}$ sampled at $200$ equally distanced points, at an excitation power below $10\,\mathrm{nW}$. The scan is performed under simultaneous microwave driving, which mixes the ground-state spin populations and thereby increases the PLE contrast by preventing from optical pumping into a dark spin state. The absolute emission frequency is obtained from the center of a Lorentzian fit to the resulting spectrum, referenced to the laser frequency recorded by the wavemeter (HighFinesse WS7).

An acceptance window of $\pm 20\,\mathrm{MHz}$ from the target emission frequency of $\nu_T = 327.231000\,\mathrm{THz}$ was preset. If the fitted $A_2$ emission frequency exceeded this acceptance window, a corrective change of the reverse bias voltage was applied, calculated from the linear tuning coefficient $\delta\nu/\delta U_{\mathrm{Bias}} = 6.30\,\mathrm{GHz}\,/\,\mathrm{V}$. This value is the common slope of the linear DC Stark shift fits of the two selected $V_{Si}$ centers (Fig. 1e). A second PLE scan then verifies that the corrected $A_2$ emission frequency lies inside the acceptance window. If it did not, the correction is repeated for up to three attempts in total. Should all three attempts fail, the run is flagged and all subsequent data is excluded from the integrated correlation histogram. In post selective analysis these correlation measurement sequences were then excluded. The two nodes were stabilized consecutively, node A first, while the other node remained idle. The procedure was executed automatically by a programmed hardware control script without user intervention. Across the $244$ verification sessions of the $628\,\mathrm{h}$ campaign, a corrective voltage change was required in $12.7\%$ of the sessions for $V_{Si,1A}$ and in $30.7\%$ for $V_{Si,2B}$, corresponding to the mean correction intervals of $19.4\,\mathrm{h}$ and $8.4\,\mathrm{h}$ reported in the main text. The stabilization sequence is shown in Fig. SI7.

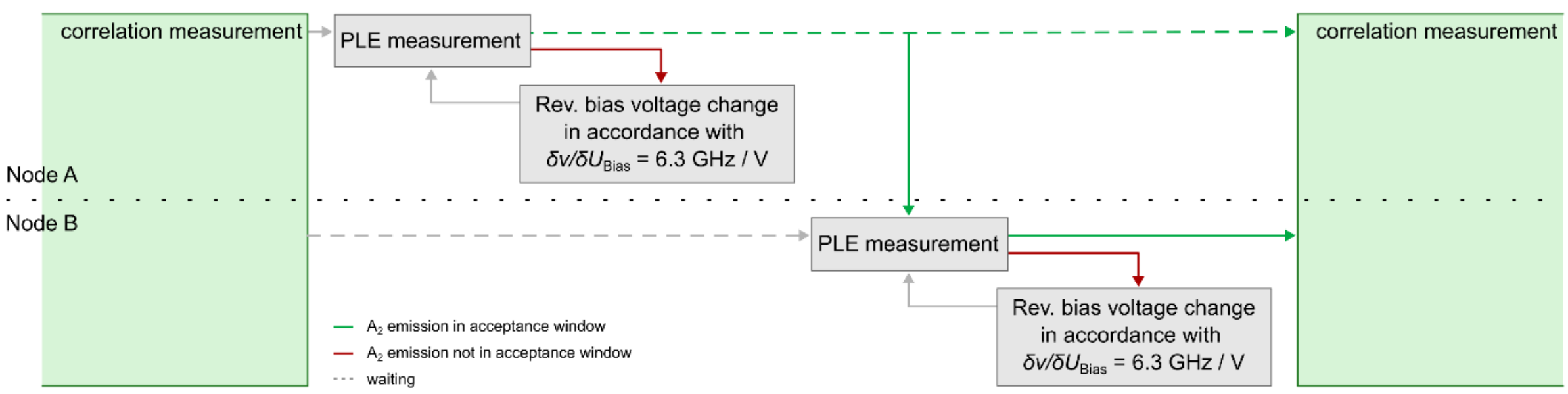

**Fig. SI7 Scheme of remote, automated control for common target emission frequency $\nu_T$.**

Timeline of the stabilization sequence, interleaved within the two-photon interference measurement, shown separately for node A (top) and node B (bottom). After every $2\,\mathrm{h}$ of active, accumulated correlation time (green blocks) the correlation measurement block is interrupted and a resonant PLE spectrum of the $A_2$ transition is recorded at node A. If the emission frequency extracted from a Lorentzian fit lies outside the acceptance window of $\pm 20\,\mathrm{MHz}$ around the common target frequency $\nu_T = 327.231000\,\mathrm{THz}$ (red arrow), a corrective change of the reverse bias voltage is applied using the linear tuning coefficient $\delta\nu/\delta U_{\mathrm{Bias}} = 6.30\,\mathrm{GHz}$ / V, and the PLE measurement is repeated. Once the emission frequency lies within the acceptance window (green arrow), the same sequence is executed for node B while node A remains idle (dotted arrows). After both nodes have been stabilized, correlation measurement resumes.